\documentclass[twocolumn,showpacs,amsmath,aps,pra,amssymb,footinbib]{revtex4-1}

\usepackage[T1]{fontenc}
\usepackage[latin9]{inputenc}
\usepackage{color}
\usepackage{float}
\usepackage{amsmath}
\usepackage{amssymb}
\usepackage{graphicx}
\usepackage{wasysym}
\usepackage[unicode=true,
 bookmarks=false,
 breaklinks=false,pdfborder={0 0 1},colorlinks=true]
 {hyperref} 
\hypersetup{
 unicode,breaklinks,a4paper=true,plainpages=false,linkcolor=blue,citecolor=blue,filecolor=black,urlcolor=blue}

\makeatletter
\usepackage{times}
\usepackage{bm}
\usepackage{epstopdf}
\usepackage{float}
\usepackage{adjustbox}

\global\long\def\bk{\mathbf{k}}

\global\long\def\br{\mathbf{r}}

\makeatother

\begin{document}

\title{Analytic solutions for the spatial character and coherence properties
of light scattered from two dipole-coupled atoms}

\author{Petra~Fersterer}

\author{R.~J.~Ballagh}

\affiliation{The Dodd-Walls Centre for Photonic and Quantum Technologies, New
Zealand}

\affiliation{Department of Physics, University of Otago, Dunedin 9016, New Zealand}
\begin{abstract}
Analytic solutions for steady-state expectation values of atomic quantities
and second order correlations are obtained for a fully quantum treatment
of two stationary dipole-coupled atoms driven in a standard geometric
configuration by a near resonant laser. Explicit expressions for the
spatial and coherence properties of the far-field scattered light
intensity are derived, valid for the full range of system parameters.
A comprehensive survey of the steady-state scattering behaviour is
given, with key features precisely characterised, including \textcolor{black}{suppression of}
scattering, and the regime in which the dipole-dipole coupling has
significant effect. A regime is also found where the incoherent scattered
light develops spatial interference fringes. We examine in detail
a decorrelation approximation that has potential application for larger
systems of atoms that are intractable in a full quantum treatment.
Finally, we introduce the concept of an effective driving field and
show that it can provide a direct and intuitive physical interpretation
of key aspects of the system behaviour. 
\end{abstract}
\maketitle

\section{Introduction \label{sec:FormalFoundation}}

Collective light scattering from a coherently driven ensemble of atoms
is a research area of long standing \cite{Lehmberg:1970tz}, but the
phenomenon remains the subject of considerable current interest e.g.
Refs. \cite{PhysRevLett.112.113603,Pellegrino:2014fj,PhysRevA.92.063822,Bromley:2016fi,Zhu:2016ds,Bettles:2016cg,Javanainen:2016wc,PhysRevA.94.013847,PhysRevA.96.053629,PhysRevA.95.053625,PhysRevA.97.053816,PhysRevA.98.013622}. In a seminal paper, Lehmberg \cite{Lehmberg:1970tz} derived
a set of operator equations to describe the response of a system of
$N$ two-state atoms coupled by vacuum radiation and driven by a monochromatic
laser, and gave general expressions for the radiation rates and spectral
properties of the scattered radiation. Lehmberg's work was motivated
by the closely related phenomenon of coherent collective spontaneous
emission, an area pioneered earlier by Dicke \cite{Dicke:1954zz},
who recognised that a sample of dipole coupled atoms could exhibit
both subradiant and superradiant emission. As a practical demonstration
of his formalism Lehmberg calculated, in a separate paper \cite{Lehmberg:1970wv},
the radiation rates and spatial intensity patterns of collective spontaneous
emission from two atoms. A major review of both the theoretical and
practical aspects of collective spontaneous emission was given by
Gross and Haroche \cite{Gross:1982js} some decades ago, but this
area too has remained of strong active interest, with experimental
milestones such as the first observation of subradiance \cite{PhysRevLett.54.1917},
observation of superradiance and subradiance from two ions \cite{PhysRevLett.76.2049},
and recently the observation of subradiance in a large dilute cold
atom gas \cite{Guerin:2016dp}.

Much of the recent focus has been on low intensity scattering from
clouds of ultracold atoms (e.g. \cite{Pellegrino:2014fj,Bromley:2016fi,Zhu:2016ds,Javanainen:2016wc,PhysRevA.96.053629,PhysRevA.97.053816},
or arrays of atoms e.g. \cite{PhysRevA.92.063822,Bettles:2016cg,PhysRevA.94.013847}).
Comparison between experimental results and theory has raised questions
of our understanding of these phenomena in certain regimes e.g. \cite{PhysRevA.97.053816}.
The role of quantum correlations between atoms is known to be important,
and their treatment requires a microscopic quantum approach. However
the exponential growth of the Hilbert space with atom number necessitates
approximate solution methods, even for systems of a few atoms. Most
theoretical treatments employ the simplifying assumption of a weak driving
field. An exception is the work of Pucci \textit{et al}.  \cite{PhysRevA.95.053625}
who have developed a large scale approximate simulation method for
a strongly driven cold gas, with a validity regime that has enabled
the role of long-range correlations to be elucidated. Physical insight
into the behaviour of the system underlies the development of these,
and future approximation methods. 

The collective scattering behaviour of two monochromatically driven
atoms is a fundamental building block for understanding the larger
scale behaviour. Theoretical results for the scattered intensity have
been presented in a number of earlier papers, but these are either
numerical solutions or analytic expressions with restricted validity
regime. For example Ku\'s and W\'odkiewicz \cite{PhysRevA.23.853} gave
an analytic expression for the temporal spectra valid for exact resonance,
small atomic separation, and large laser intensity. Rudolph \textit{et
al.} \textcolor{black}{\cite{Rudolph:1995cn}} gave detailed results
for the spatial pattern and spectra of the scattered radiation using
a numerical implementation of an eigenmode approach. Wong \textit{et
	al.} \cite{Wong:1997fc}
used a quantum Monte Carlo method to study the spatial interference
pattern and polarization of the intensity scattered from two closely
separated $j=1/2\longleftrightarrow j=1/2$ atoms. 

In this paper we obtain compact analytic solutions for the coherence
and spatial properties of the intensity scattered from two $j=0\longleftrightarrow j=1$
atoms driven in a typical geometry by a monochromatic laser. Our solutions
are valid for a full range of the system parameters and permit precise
analytic characterisation of the key features of the behaviour that
can occur, including \textcolor{black}{suppression of scattering at very small interatomic separation}. In addition, with the
motivation of developing a tractable approximate solution method for
larger systems, we investigate in detail an approximation scheme for
our system which sharply reduces the number of system equations. This
approximation method, which is based on a decorrelation procedure, is shown to be very accurate over a wide range of the system parameters. 
Furthermore this approximation provides a physical interpretation of key aspects of
the system behaviour in familiar electromagnetic terms, and for example
leads to an intuitive explanation of \textcolor{black}{the modulation of scattered intensity with interatomic distance}.  

The paper is organised as follows. In Section \ref{sec:Formalism}
we outline the formalism used, and the derivation of the quantum Langevin
equations for the atomic operators. Choosing linear polarisation for
the laser, and a specific geometrical configuration, the atoms reduce
to effectively two-state. We present the equations for the ensemble
averages of the atomic quantities and correlations required to construct
the scattered intensity, and an analytic solution for those quantities.
In Section \ref{sec:Results_Perp} we provide a compact and comprehensive
survey of the behaviour of the steady state scattered intensity over
the entire parameter regime. We obtain analytic characterisations
for the features observed, including the spatial interference fringes
in the incoherent intensity. In Section \ref{subsec:Decorrelation-Approximation}
we introduce the decorrelation approximation and provide a quantitative
 analysis of its validity regime. We also introduce the concept of
the effective field, and show how it can be used to give a physical
interpretation of certain key features of the behaviour of the scattered
field. Finally, in Section \ref{sec:Other-Geometries} we consider
a second archetypal geometric configuration, and with a selection
of numerical results, demonstrate the utility of the effective field
concept in explaining their prominent features. 

\section{Formalism \label{sec:Formalism}}

\subsection{Overview\label{subsec:Overview}}

\begin{figure}
\includegraphics[width=0.48\textwidth]{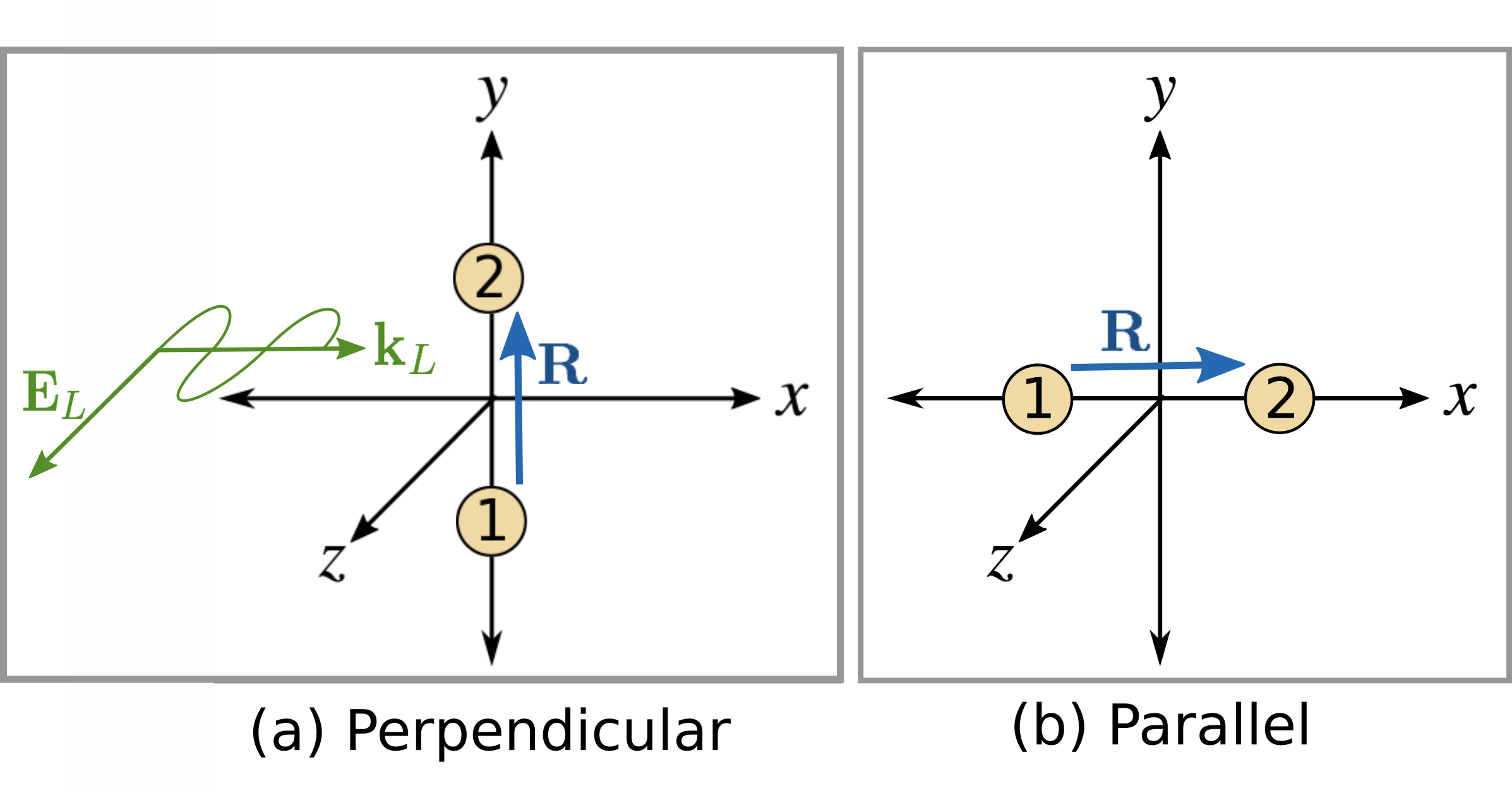}

\caption{\label{fig:System} \textcolor{black}{System diagram indicating the directions of the wavevector and polarisation of the laser field and the perpendicular and   parallel  atomic configurations.}}
\end{figure}

We consider two identical atoms, each with an optical dipole transition
between a lower level $l$ and upper level $u$ with angular momenta
$j_{l}=0$ and $j_{u}=1$ respectively. The atoms, which we assume
to be stationary, interact with an external cw single mode laser and
the vacuum radiation field. The laser field is a coherent state and
can be treated as a classical field \cite{CT:PhotsAtoms}, which
we choose to be linearly polarised with wave vector $\bk_{L}=k_{L}\mathbf{\hat{x}}$
and amplitude of the electric displacement $\mathbf{D}^{\text{ext}}=\epsilon_{0}E_{L}\mathbf{\hat{z}} $ \textcolor{black}{(see  Fig. \ref{fig:System}).}

The Power-Zienau-Wooley formulation of Quantum Electrodynamics, described
in depth in the text by Cohen-Tannoudji \textit{et al. } \cite{CT:PhotsAtoms},
is the most convenient for this problem. This approach, in which the
quantised field is the transverse electric displacement, $\mathbf{D}_{\perp}(\br)$,
has the important advantage that the interaction between separated
atoms is entirely due to the quantised fields and has no longitudinal
(Coulomb) field contribution \footnote{There is also a ``contact'' collisional term arising from the polarisation
	of the atoms, see the final term in Eq.(\ref{eq:HamDef})}. Morice \textit{et al.} \cite{Morice:1995gf} have used this formalism
to derive the evolution equations for a gas of identical $j=0\leftrightarrow j=1$
atoms driven by a weak external laser field, and we will adopt their
method, but derive equations valid for arbitrary laser intensity.
In the following, we discuss some key points in the derivation but
present only equations necessary for our current purpose. More details
can be found in ref. \cite{Morice:1995gf}, and in the appendix of
this paper where we present the full set of evolution equations for
arbitrary geometry.

The Hamiltonian for the system in the dipole approximation is 
\begin{eqnarray}
H & = & \sum_{j=1}^{2}\left[\frac{\mathbf{p}_{j}^{2}}{2m}+\left(\omega_{A}^{0}+\delta\omega_{A}\right)\mathbf{1}_{u}^{\left(j\right)}\right]\nonumber \\
 &  & +\int_{|\bk|<k_{M}}d^{3}k\sum_{\boldsymbol{\varepsilon}\perp\bk}\hbar\omega_{k}\left[a_{\bk\boldsymbol{\varepsilon}}^{\dagger}a_{\bk\boldsymbol{\varepsilon}}+\frac{1}{2}\right]\nonumber \\
 &  & -\sum_{j=1}^{2}\frac{1}{\epsilon_{0}}\mathbf{d}_{j}\cdot\left[\mathbf{D}_{\perp}\left(\mathbf{r}_{j}\right)+\mathbf{D}^{{\rm ext}}\left(\mathbf{r}_{j}\right)\right]\nonumber \\
 &  & +\frac{1}{\epsilon_{0}}\mathbf{d}_{1}\cdot\mathbf{d}_{2}\delta\left(\mathbf{r}_{1}-\mathbf{r}_{2}\right),\label{eq:HamDef}
\end{eqnarray}
where $\mathbf{r}_{j},\mathbf{p}_{j},$ are the position and momentum
operators for the center of mass of the $j^{th}$ atom, $\mathbf{d}_{j}$
is the dipole operator and $\mathbf{1}_{u}^{\left(j\right)}$ is the
unit operator in the upper level subspace of the $j^{th}$ atom. The
transition has a ``bare'' frequency of $\omega_{A}^{0}$, while
the quantity $\delta\omega_{A}$ is the net shift of the bare transition
frequency due to the dipole self energy. The second term on the RHS
of Eq.(\ref{eq:HamDef}) is the free Hamiltonian for the displacement
field, with $a_{{\bf k}\boldsymbol{\varepsilon}}$ and $a_{{\bf k}\boldsymbol{\varepsilon}}^{\dagger}$
the annihilation and creation operators for mode $\left({\bf k},\boldsymbol{\varepsilon}\right)$.
The integral has a cutoff at $k_{M}$ which is required for self consistency
in a nonrelativistic treatment of the atoms \cite{CT:PhotsAtoms}.
The third term in Eq.(\ref{eq:HamDef}) represents the dipole interaction
of the atoms with the laser field ($\mathbf{D}^{{\rm ext}}$) and
the internal radiation field ($\mathbf{D}_{\perp}$), which is transverse.
Morice \textit{et al. }write the field $\mathbf{D}\left(\br\right)/\epsilon_{0}$
as $\mathbf{E}\left(\br\right)$ and call it the ``electric displacement
vector, up to a factor of $\epsilon_{0}$'', and we shall henceforth
follow their practise, and define 
\begin{equation}
\mathbf{E}\left(\br\right)\equiv\mathbf{D}_{\perp}\left(\br\right)/\epsilon_{0}\thinspace;\ \mathbf{E}_{L}\left(\br\right)\equiv\mathbf{D}^{{\rm ext}}\left(\br\right)/\epsilon_{0}\label{eq:E_defn_disp}
\end{equation}
We note that $\mathbf{D}\left(\br\right)/\epsilon_{0}$ is identical
to the electric field away from the atoms (i.e. in vacuum). The final
term in Eq.(\ref{eq:HamDef}) represents contact interaction between
atoms, which we will henceforth ignore, assuming the atoms are always
separated. We denote the spatial separation of the atoms as ${\bf R\equiv\br}_{2}-\br$$_{1}$,
and in all that follows we will position the atoms symmetrically around
the origin of the coordinate axes so that $\br$$_{1}=-\frac{{\bf R}}{2}$
and $\br$$_{2}=\frac{{\bf R}}{2}$. The equations of motion for the
electric field and the atomic operators are derived by the quantum
Heisenberg-Langevin method, which is described in depth for the case
of a single two-state atom in the text by Cohen-Tannoudji \textit{et al. } \cite{CT:AtomPhotInt}. Adaptions to the multi-atom case are
briefly presented by Morice \textit{et al. }\cite{Morice:1995gf}.
In this paper we express the dipole operator for the $j^{th}$ atom
as 
\begin{equation}
\mathbf{\mathbf{\mathbf{d}}}_{j}=\mathcal{D}\sum_{q=-1,0,1}\left[d_{+}^{q,j}e^{i\omega_{L}t}+d_{-}^{q,j}e^{-i\omega_{L}t}\right]\hat{\mathbf{e}}_{q}^{*}\label{eq:DipoleDefn}
\end{equation}
where expansion on the standard unit spherical vectors $\hat{\mathbf{e}}_{q}^{*}$
is chosen for convenience in applying selection rules and symmetries
\cite{Edmonds}. Here $\mathcal{D}\equiv(j_{u}||\mathbf{d}||j_{l})/\sqrt{3}$
and the reduced dipole matrix element \cite{Edmonds} $(j_{u}||\mathbf{d}||j_{l})$
is chosen to be positive \cite{Ducloy:1973gq}. The slowly varying
operators $d_{+}^{q,j}$ and $d_{-}^{q,j}$ can be expressed in terms
of irreducible spherical tensors (see Appendix), and correspond respectively
to raising and lowering operators with their $t=0$ nonzero matrix
elements given by $\langle1q|d_{+}^{q}|00\rangle=1$ and $\langle00|d_{-}^{q}|1-q\rangle=\left(-\right)^{q}$.
The first step in the derivation of the Langevin equations is to obtain
an expression for the electric field operator by formally integrating
its equation of motion and applying well understood approximations
(including the Markov approximation, see ref \cite{Morice:1995gf})
to give

\begin{eqnarray}
\mathbf{E}\left(\mathbf{r},t\right) & = & \frac{\hbar}{\mathcal{D}}\sum_{j}\left[e^{i\omega_{L}t}g^{\dagger}(\mathbf{R}_{j})\boldsymbol{d}_{+}^{(j)}\left(t\right)+e^{-i\omega_{L}t}g(\mathbf{R}_{j})\boldsymbol{d}_{-}^{(j)}\left(t\right)\right]\nonumber \\
 &  & +\mathbf{f}\left(\br,t\right),\label{eq:Escatt_G}
\end{eqnarray}
Here $\mathbf{R}_{j}\equiv{\bf r}-\mathbf{r}_{j}$ , $\boldsymbol{d}_{\pm}^{(j)}\left(t\right)=\sum_{q}d_{\pm}^{q,j}\hat{\mathbf{e}}_{q}^{*}$,
and $\mathbf{f}\left(\mathbf{r},t\right)$ is the quantum noise component
of the field, which is given in full form in the Appendix. The quantity
$g(\mathbf{R}_{j})$ is a matrix given in spherical coordinates by
\begin{eqnarray}
g_{\alpha\beta} (\mathbf{R}_{j})& &=  \frac{3\gamma}{4}e^{ik_{L}R_{j}}\label{eq:arbgdef}\\
 & & \bigg[\delta_{\alpha\beta}\left(-\frac{1}{(k_{L}R_{j})^{3}}+\frac{i}{(k_{L}R_{j})^{2}}+\frac{1}{k_{L}R_{j}}\right)\nonumber \\
 & & -\frac{R_{j,\alpha}R_{j,\beta}^{*}}{R_{j}^{2}}\left(-\frac{3}{(k_{L}R_{j})^{3}}+\frac{3i}{(k_{L}R_{j})^{2}}+\frac{1}{k_{L}R_{j}}\right)\bigg],\nonumber
\end{eqnarray}
where $\gamma$ is the Einstein A coefficient for the transition,
$R_{j}=|\mathbf{R}_{j}|$ and $R_{j,\alpha}=\mathbf{R}_{j}\cdot\hat{\mathbf{e}}_{\alpha}$.
This expression, valid for $\mathbf{R}_{j}\ne0$, provides the familiar
spatial dependence of the electric field scattered from an oscillating
dipole e.g. see refs. \cite{Lehmberg:1970tz,Milonni:1974tf,Jackson:1975}.
We have omitted $\delta\left(\mathbf{R}\right)$ terms from the RHS
of Eq.(\ref{eq:arbgdef}), which are required to obtain the correct
$\lim_{\mathbf{R}_{j}\rightarrow0}g(\mathbf{R}_{j})$, and are retained
where necessary in our derivation to describe self-field effects.
These give rise to radiative damping, a radiative correction that
changes $\omega_{A}^{0}$ to a true resonance frequency $\omega_{A}$,
and a term that cancels the dipole self energy $\delta\omega_{A}$
\cite{Morice:1995gf,Dalibard:1982ig}. The $\lim_{\mathbf{R}_{j}\rightarrow0}g(\mathbf{R}_{j})$
would also be needed to describe the transverse field interaction
between atoms in contact, but we have excluded this possibility in
our model, and thus Eq.(\ref{eq:arbgdef}) is appropriate for our
purposes. \textcolor{black}{Eq.(\ref{eq:arbgdef}) includes the familiar
near field ($R^{-3}$ and $R^{-2}$) and far field ($R^{-1}$) terms
, but for convenience in this paper we shall refer to all of these
parts of the field together as the }\textit{\textcolor{black}{scattered
field}}\textcolor{black}{. Where necessary to avoid ambiguity we will
use the descriptor }\textit{\textcolor{black}{far-field}}\textcolor{black}{{}
to designate the scattered electric intensity that arises from the
($R^{-1}$) terms.}

Expression (\ref{eq:Escatt_G}) for $\mathbf{E}\left(\br\right)$
is now substituted wherever $\mathbf{E}$ appears in the equations
for the atomic quantities, leading to the quantum Langevin equations.
It is worth noting, as first recognised by Milonni and Knight \cite{Milonni:1974tf},
that in treating the resonant interaction between two atoms, it is
essential that the full dipole interaction in the Hamiltonian is retained
(i.e. including the non-energy conserving terms that are neglected
in the usual rotating wave approximation (RWA)), in order that the
correct atomic shifts and retardation times are obtained. Instead,
a RWA is made on the final quantum Langevin equations, which also
ensures we obtain the correct correspondence to the classical version
of the problem.\\

\subsection{Equations of motion }

The observable of interest in this paper is the mean scattered intensity,
which is proportional to $\langle\mathbf{E}^{\left(-\right)}\left(\br\right)\mathbf{E}^{\left(+\right)}\left(\br\right)\rangle$
where $\mathbf{E}^{\left(+\right)}$ and $\mathbf{E}^{\left(-\right)}$
are the positive and negative frequency components of the electric
field. We see from Eq.(\ref{eq:Escatt_G}) that we therefore need
solutions for mean atomic quantities such as $\langle d_{+}^{q,j}\rangle$
, $\langle d_{+}^{q,j}d_{-}^{q',i}\rangle$ and others, and the evolution
equations required for these quantities are obtained by taking mean
values of the operator equations presented in the Appendix. The mean
in the Heisenberg picture is taken with an initial system state of
the form $|\alpha\rangle|{\rm vac}\rangle$ where $|\alpha\rangle$
is some choice of internal states for the two atoms and $|{\rm vac}\rangle$
is the vacuum state of the radiation field. Thus the noise terms in
operator equations, where the field operators are normally ordered,
disappear in the mean equations.

The most important features of our problem are present for the specific
geometry where ${\bf R}$ is parallel to the $y$ axis, which we will
call \emph{perpendicular configuration}  \textcolor{black}{(see Fig. \ref{fig:System})} and we will concentrate our
study on this case. One other simple configuration will be considered
briefly later in the paper. In each of the two configurations we consider,
the incident and scattered light interacting with the atoms is polarised
along $\mathbf{\hat{z}}$, and hence only the lower atomic state ($|00\rangle$)
and the $m=0$ upper state ($|10\rangle$) of each atom participate
in the interactions. The atoms are each reduced to the familiar two-state
case, the only dipole operators needed are $d_{\pm}^{0,j}$, and the
number of equations required reduces from $255$ to $15$. It is appropriate
for these effectively two-state atoms to use simpler atomic notation,
so that we write $d_{\pm}^{(j)}$ for $d_{\pm}^{0,j}$ and $n_{u}^{(j)}$
$(n_{l}^{(j)})$ for the operator for the upper (lower) state population.
From Eqs.(\ref{eq:Tlu_full_ex}) to (\ref{eq:TuuTuu_full_ex}) we
obtain the following equations for the first ten atomic mean quantities 
\begin{widetext}
\begin{align}
-i\frac{d}{dt}\langle d_{-}^{(i)}\rangle= & \left(\Delta+i\frac{\gamma}{2}\right)\langle d_{-}^{(i)}\rangle+\frac{\Omega e^{i\mathbf{k}_{L}\cdot{\bf r}_{i}}}{2}\left(1-2\langle n_{u}^{(i)}\rangle\right)+G(\mathbf{R})\left(\langle d_{-}^{(j)}\rangle-2\langle d_{-}^{(j)}n_{u}^{(i)}\rangle\right),\label{eq:dlu}
\end{align}

\begin{align}
-i\frac{d}{dt}\langle n_{u}^{(i)}\rangle= & i\gamma\langle n_{u}^{(i)}\rangle+\frac{\Omega}{2}\left(e^{i\mathbf{k}\cdot{\bf r}_{i}}\langle d_{+}^{(i)}\rangle-e^{-i\mathbf{k}_{L}\cdot{\bf r}_{i}}\langle d_{-}^{(i)}\rangle\right)+G(\mathbf{R})\langle d_{+}^{(i)}d_{-}^{(j)}\rangle-G^{*}(\mathbf{R})\langle d_{-}^{(i)}d_{+}^{(j)}\rangle,\label{eq:nu}
\end{align}

\begin{align}
-i\frac{d}{dt}\left\langle d_{-}^{(i)}d_{+}^{(j)}\right\rangle = & i\gamma\left\langle d_{-}^{(i)}d_{+}^{(j)}\right\rangle +G(\mathbf{R})\langle n_{u}^{(j)}\rangle-G^{*}(\mathbf{R})\langle n_{u}^{(i)}\rangle-2\left(G(\mathbf{R})-G^{*}(\mathbf{R})\right)\langle n_{u}^{(i)}n_{u}^{(j)}\rangle\nonumber \\
 & +\frac{\Omega}{2}\left(e^{i\mathbf{k}_{L}\cdot{\bf r}_{i}}\langle d_{+}^{(j)}\rangle-2e^{i\mathbf{k}_{L}\cdot{\bf r}_{i}}\langle n_{u}^{(i)}d_{+}^{(j)}\rangle-e^{-i\mathbf{k}_{L}\cdot{\bf r}_{j}}\langle d_{-}^{(i)}\rangle+2e^{-i\mathbf{k}_{L}\cdot{\bf r}_{j}}\langle d_{-}^{(i)}n_{u}^{(j)}\rangle\right),\label{eq:dludul}
\end{align}

\begin{align}
-i\frac{d}{dt}\langle n_{u}^{(i)}d_{-}^{(j)}\rangle= & \left(\Delta+i\frac{3\gamma}{2}\right)\langle n_{u}^{(i)}d_{-}^{(j)}\rangle-G^{*}(\mathbf{R})\langle d_{-}^{(i)}n_{u}^{(j)}\rangle\nonumber \\
 & +\frac{\Omega}{2}\left(e^{i\mathbf{k}_{L}\cdot{\bf r}_{j}}\left(\langle n_{u}^{(i)}\rangle-2\langle n_{u}^{(i)}n_{u}^{(j)}\rangle\right)+e^{i\mathbf{k}_{L}\cdot{\bf r}_{i}}\langle d_{+}^{(i)}d_{-}^{(j)}\rangle-e^{-i\mathbf{k}_{L}\cdot{\bf r}_{i}}\langle d_{-}^{(i)}d_{-}^{(j)}\rangle\right),\label{eq:nudlu}
\end{align}

\begin{align}
-i\frac{d}{dt}\langle d_{-}^{(1)}d_{-}^{(2)}\rangle= & \left(2\Delta+i\gamma\right)\langle d_{-}^{(1)}d_{-}^{(2)}\rangle+\frac{\Omega}{2}\left(e^{i\mathbf{k}_{L}\cdot{\bf r}_1}\langle d_{-}^{(2)}\rangle+e^{i\mathbf{k}_{L}\cdot{\bf r}_2}\langle d_{-}^{(1)}\rangle-2e^{i\mathbf{k}_{L}\cdot{\bf r}_1}\langle n_{u}^{(1)}d_{-}^{(2)}\rangle-2e^{i\mathbf{k}_{L}\cdot{\bf r}_2}\langle d_{-}^{(1)}n_{u}^{(2)}\rangle\right),\label{dludlu}
\end{align}

\begin{align}
-i\frac{d}{dt}\langle n_{u}^{(1)}n_{u}^{(2)}\rangle= & 2i\gamma\langle n_{u}^{(1)}n_{u}^{(2)}\rangle+\frac{\Omega}{2}\left(e^{i\mathbf{k}_{L}\cdot{\bf r}_1}\langle d_{+}^{(1)}n_{u}^{(2)}\rangle+e^{i\mathbf{k}_{L}\cdot{\bf r}_2}\langle n_{u}^{(1)}d_{+}^{(2)}\rangle-e^{-i\mathbf{k}_{L}\cdot{\bf r}_1}\langle d_{-}^{(1)}n_{u}^{(2)}\rangle-e^{-i\mathbf{k}_{L}\cdot{\bf r}_2}\langle n_{u}^{(1)}d_{-}^{(2)}\rangle\right).\label{eq:nunu}
\end{align}
\end{widetext}

The remaining five equations are easily found by noting $d_{+}^{(j)}=\left(d_{-}^{(j)}\right)^{\dagger}$
which gives $\langle d_{+}^{(i)}\rangle=\langle d_{-}^{(i)}\rangle^{*}$,
$\langle n_{u}^{(i)}d_{+}^{(j)}\rangle=\langle n_{u}^{(i)}d_{-}^{(j)}\rangle^{*}$
and $\langle d_{+}^{(1)}d_{+}^{(2)}\rangle=\langle d_{-}^{(1)}d_{-}^{(2)}\rangle^{*}$.
In these equations $i\neq j$, $\Delta\equiv\omega_{L}-\omega_{A}$
is the laser detuning, $\Omega=\mathcal{D}E_{L}/\hbar$ is the Rabi
frequency, \textcolor{black}{and $G(\mathbf{R})\equiv g_{00}(\mathbf{R})$}
(note that $R_{0}=R_{z}$). Equations (\ref{eq:dlu})-(\ref{eq:nunu})
are formally equivalent to those given by Lehmberg \cite{Lehmberg:1970tz}
in his seminal paper on collective light scattering, and can be mapped
directly to those given by Rudolf \textit{\textcolor{black}{et al.} }
\cite{Rudolph:1995cn}.

\subsubsection{Analytic solution for perpendicular configuration\label{subsec:Analytic-solution-for}}

In the case of perpendicular configuration ($\mathbf{R}$ parallel
to $y$-axis), the factors $e^{i\mathbf{k}_{L}\cdot{\bf r}_{i}},e^{i\mathbf{k}_{L}\cdot{\bf R}}\rightarrow1$,
simplifying the equations of motion in the previous section and allowing
the following steady state solution to be obtained \footnote{ Note these analytic solutions also hold when $\mathbf{R}\parallel\mathbf{\hat{z}}$},

\begin{align}
\langle d_{-}^{(1)}\rangle=\langle d_{-}^{(2)}\rangle= & -\frac{\Omega}{A}(2\Delta-i\gamma)\label{eq:dlu_analytic}\\
 & \left[2\Omega^{2}+\left(2\Delta+i\gamma\right)\left(2\Delta-i\gamma+2G^{*}(\mathbf{R)}\right)\right],\nonumber 
\end{align}

\begin{equation}
\langle n_{u}^{(1)}\rangle=\langle n_{u}^{(2)}\rangle=\frac{\Omega^{2}}{A}(4\Delta^{2}+2\Omega^{2}+\gamma^{2}),\label{eq:nu_analytic}
\end{equation}

\begin{equation}
\langle d_{+}^{(1)}d_{-}^{(2)}\rangle=\langle d_{+}^{(2)}d_{-}^{(1)}\rangle=\frac{\Omega^{2}}{A}\left(4\Delta^{2}+\gamma^{2}\right),\label{eq:dlu_dul_analytic}
\end{equation}

\begin{equation}
\langle n_{u}^{(1)}d_{-}^{(2)}\rangle=\langle n_{u}^{(2)}d_{-}^{(1)}\rangle=-\frac{\Omega^{3}}{A}(2\Delta-i\gamma),\label{eq:nu_dm_corr}
\end{equation}

\begin{equation}
\langle d_{-}^{(1)}d_{-}^{(2)}\rangle=\frac{\Omega^{2}}{A}(2\Delta-i\gamma)(2\Delta-i\gamma+2G^{*}(\mathbf{R})),\label{eq:dm_dm_corr}
\end{equation}

\begin{equation}
\langle n_{u}^{(1)}n_{u}^{(2)}\rangle=\frac{\Omega^{4}}{A},\label{eq:nu_nu_corr}
\end{equation}
where

\begin{align}
A= & \left(\gamma^{2}+4\Delta^{2}\right)\left[(2\text{\ensuremath{G_{i}}(\ensuremath{\mathbf{R}})}+\gamma)^{2}+4(\text{\ensuremath{G_{r}}(\ensuremath{\mathbf{R)}}}+\Delta)^{2}+4\Omega^{2}\right]\nonumber \\
 & +4\Omega^{4},
\end{align}
and $G_{r}(\mathbf{R})$ and $G_{i}(\mathbf{R})$ are the real and
imaginary parts of $G(\mathbf{R}).$

\subsection{Observables \label{subsec:Observables}}

In this paper, our primary interest is in the spatial distribution
of the steady-state far-field scattered intensity, which in the far-field
approximation ($r\gg\lambda_{L},R$) is given by

\begin{align}
I(\mathbf{r})= & \frac{P_{0}}{r^{2}}\left(1-\frac{r_{z}^{2}}{r^{2}}\right)\label{eq:Intensity_dist}\\
&\left(\langle n_{u}^{(1)}\rangle+\langle n_{u}^{(2)}\rangle+2\Re\left(e^{-ik_{L}\hat{\mathbf{r}}\cdot\mathbf{R}}\langle d_{+}^{(1)}d_{-}^{(2)}\rangle\right)\right),\nonumber
\end{align}
with $P_{0}=3\gamma\hbar\omega_{L}/8\pi$. The coherent part of the
far-field intensity, which is proportional to $\langle\mathbf{E}^{\left(-\right)}\left(\br\right)\rangle\langle\mathbf{E}^{\left(+\right)}\left(\br\right)\rangle$,
is given by

\begin{align}
I&_{{\rm coh}}  (\mathbf{r})=\frac{P_{0}}{r^{2}}\left(1-\frac{r_{z}^{2}}{r^{2}}\right)\label{eq:Intensity_coh}\\
 & \left(\langle d_{+}^{(1)}\rangle\langle d_{-}^{(1)}\rangle+\langle d_{+}^{(2)}\rangle\langle d_{-}^{(2)}\rangle+2\Re\left(e^{-ik_{L}\hat{\mathbf{r}}\cdot\mathbf{R}}\langle d_{+}^{(1)}\rangle\langle d_{-}^{(2)}\rangle\right)\right),\nonumber 
\end{align}
and the incoherent scattering by $I_{{\rm inc}}(\mathbf{r})=I(\mathbf{r})-I_{{\rm coh}}(\mathbf{r})$.
The far-field radiation $I(\mathbf{r})$ forms a pattern of interference
fringes characterised by the well known $g^{\left(1\right)}$ correlation
factor \cite{Gardiner:1991wq}, which here simplifies to

\begin{equation}
g^{\left(1\right)}(\mathbf{r}_{1};\br_{2})=\frac{\left\langle d_{+}^{(2)}d_{-}^{(1)}\right\rangle }{\sqrt{\langle n_{u}^{(1)}\rangle\langle n_{u}^{(2)}\rangle}}.\label{eq:g1_def}
\end{equation}
The visibility of the fringes in the $x-y$ plane is

\begin{equation}
V\equiv\frac{I_{{\rm max}}-I_{{\rm min}}}{I_{{\rm max}}+I_{{\rm min}}}=\frac{2|g^{\left(1\right)}(\mathbf{r}_{1};\br_{2})|}{\sqrt{\frac{\langle n_{u}^{(1)}\rangle}{\langle n_{u}^{(2)}\rangle}}+\sqrt{\frac{\langle n_{u}^{(2)}\rangle}{\langle n_{u}^{(1)}\rangle}}},\label{eq:FringeVis}
\end{equation}
where the final equality in Eq.(\ref{eq:FringeVis}) applies when
$R\geq\lambda_{L}/2$.

\subsubsection{\textcolor{black}{Scattered Power}}

The total power scattered into the far field is obtained by integrating
the far-field intensity over all angles, which gives 
\begin{align}
P_{\text{scatt}}& =\int d\mathbf{\Omega_{r}}I(\mathbf{r})\label{eq:Pscatt}\\
&  = \frac{2\hbar\omega}{A}\left((\gamma+2G_{i}(\mathbf{r}))(\gamma^{2}+4\Delta^{2})\Omega^{2}+2\gamma\Omega^{4}\right).\nonumber
\end{align}
The power absorbed from the laser by the atoms is given by 

\begin{align}
P_{\text{\text{abs}}} & =-\omega_{L}\hbar\Omega\Im(\langle d_{+}^{(1)}\rangle+\langle d_{+}^{(2)}\rangle)\label{eq:Pabs}
\end{align}
and it is easy to show that $P_{\text{scatt}}=P_{\text{abs}}$, as
expected. 

\subsubsection{Scattering from two uncoupled atoms}

\textcolor{black}{It will be useful when discussing the results for
the scattering from two dipole-coupled atoms, to compare with the
scattering from two non-dipole-coupled atoms, driven by the same laser
field. Although the latter is an artificial model, it will allow us
to identify the features of the scattering that are due to the coupling.
Results for scattering from a single laser-driven atom were derived
many years ago by Mollow} \cite{Mollow:1969wc}\textcolor{black}{,
and for the convenience of the reader we present }his results in our
current notation for the upper state population $\left(\langle n_{u}\rangle_{M}\right)$,
and the lowering component of the dipole $\left(\langle d_{-}\rangle_{M}\right)$

\begin{equation}
\langle n_{u}\rangle_{M}=\frac{\frac{1}{4}\Omega^{2}}{\Delta^{2}+\left(\frac{\gamma}{2}\right)^{2}+\frac{1}{2}\Omega^{2}},\label{eq:nu_Mollow}
\end{equation}

\begin{equation}
\langle d_{-}\rangle_{M}=\frac{i\frac{\gamma}{2}-\Delta}{\Delta^{2}+\left(\frac{\gamma}{2}\right)^{2}+\frac{1}{2}\Omega{}^{2}}\frac{\Omega}{2}.\label{eq:dlu_Mollow}
\end{equation}
For this single atom, the coherent fraction of the scattered intensity
is given by

\begin{align}
f_{{\rm coh}}^{\text{M}} & \equiv\frac{I_{{\rm coh}}}{I}=\frac{\langle\mathbf{E}^{\left(-\right)}\rangle\cdot\langle\mathbf{E}^{\left(+\right)}\rangle}{\langle\mathbf{E}^{\left(-\right)}\cdot\mathbf{E}^{\left(+\right)}\rangle}=\frac{|\langle d_{-}\rangle_{M}|^{2}}{\langle n_{u}\rangle_{M}}\nonumber \\
 & =\frac{4\Delta^{2}+\gamma^{2}}{2\Omega{}^{2}+4\Delta^{2}+\gamma^{2}},\label{eq:fcohMollow}
\end{align}
and the incoherent fraction $f_{{\rm inc}}^{M}=1-f_{{\rm coh}}^{M}$.
It is easily shown from Eqs.(\ref{eq:dlu_analytic}) and (\ref{eq:nu_analytic})
that when $G(\mathbf{R})\rightarrow0$ (i.e. when the dipole coupling
is put to zero), $\langle n_{u}^{(i)}\rangle\rightarrow\langle n_{u}^{(i)}\rangle_{M},$
$\langle d_{-}^{(i)}\rangle\rightarrow\langle d_{-}^{(i)}\rangle_{M}$,
and furthermore each of the correlations Eqs.(\ref{eq:dlu_dul_analytic})-(\ref{eq:nu_nu_corr})
factor into Mollow results, e.g. $\langle d_{-}^{(1)}d_{+}^{(2)}\rangle\rightarrow\langle d_{-}^{(1)}\rangle_{M}\langle d_{+}^{(2)}\rangle_{M}$.
Thus the spatial intensity distribution of light scattered from two
uncoupled atoms (which we denote $I^{{\rm uc}}(\mathbf{r})$) is obtained
from Eq.(\ref{eq:Intensity_dist}) to be
\begin{align}
I^{{\rm uc}} & (\mathbf{r})=\frac{P_{0}}{r^{2}}\left(1-\frac{r_{z}^{2}}{r^{2}}\right)\label{eq:ItwoMollow}\\
 & \left(\langle n_{u}^{(1)}\rangle_{M}+\langle n_{u}^{(2)}\rangle_{M}+2\Re\left(e^{-ik_{L}\hat{\mathbf{r}}\cdot\mathbf{R}}\langle d_{+}^{(1)}\rangle_{M}\langle d_{-}^{(2)}\rangle_{M}\right)\right).\nonumber 
\end{align}
$I^{{\rm uc}}(\mathbf{r})$ has both coherent and incoherent components,
and the visibility of the fringes in the perpendicular configuration
is 
\[
V^{{\rm uc}}=\frac{2\langle d_{+}^{(1)}\rangle_{M}\langle d_{-}^{(2)}\rangle_{M}}{\langle n_{u}^{(1)}\rangle_{M}+\langle n_{u}^{(2)}\rangle_{M}}=f_{{\rm coh}}^{\text{M}}.
\]
In the forward direction\textcolor{black}{{} $I^{{\rm uc}}(\mathbf{r})$
is given by 
\begin{equation}
I_{{\rm fwd}}^{{\rm uc}}(r)=\frac{P_{0}}{r^{2}}\left[\frac{4\Omega^{2}\left(\Omega^{2}+4\Delta^{2}+\gamma^{2}\right)}{\left(2\Omega{}^{2}+4\Delta^{2}+\gamma^{2}\right)^{2}}\right],\label{eq:Ifwd_uc}
\end{equation}
of which the incoherent fraction is 
\begin{equation}
f_{{\rm fwd,inc}}^{{\rm uc}}=\frac{\Omega^{2}}{\Omega{}^{2}+4\Delta^{2}+\gamma^{2}}.\label{eq:finc_uc_fwd}
\end{equation}
We note finally that the total power scattered by the two uncoupled
atoms is $2\gamma\hbar\omega\langle n_{u}\rangle_{M}$.}

\section{Results: Perpendicular Configuration \label{sec:Results_Perp} }

The perpendicular configuration is the main focus of this paper, and
from the analytic solutions Eqs.(\ref{eq:nu_analytic}) and (\ref{eq:dlu_dul_analytic})
we see in this case $\langle n_{u}^{(1)}\rangle=\langle n_{u}^{(2)}\rangle$,
$\langle d_{+}^{(1)}d_{-}^{(2)}\rangle$ is real, and $g^{\left(1\right)}(\mathbf{r}_{1};\br_{2})$
has no $\br_{1}$ or $\br_{2}$ dependence. We will henceforth denote
it by $g_{\perp}^{\left(1\right)}$, and it has the simple form 
\[
g_{\perp}^{\left(1\right)}=g^{\left(1\right)}(\mathbf{r}_{1};\br_{2})=\frac{4\Delta^{2}+\gamma^{2}}{2\Omega{}^{2}+4\Delta^{2}+\gamma^{2}}
\]
which is identical to $f_{{\rm coh}}^{\text{M}}.$ The steady state
value of $I(\mathbf{r})$ can now be expressed\textcolor{black}{{} as
\begin{equation}
I(\mathbf{r})=\left(1-\frac{r_{z}^{2}}{r^{2}}\right)\left[\frac{1+g_{\perp}^{\left(1\right)}\cos\left(-k_{L}\hat{\mathbf{r}}\cdot\mathbf{R}\right)}{1+g_{\perp}^{\left(1\right)}}\right]I_{{\rm fwd}}(r)\label{eq:I3D_perp_config}
\end{equation}
where $I_{{\rm fwd}}(r)$ is the far-field intensity at a distance
$r$ in the forward direction (i.e. on the }\textcolor{black}{\emph{x}}\textcolor{black}{-axis), }

\begin{eqnarray}
I_{{\rm fwd}}(r) & = & \frac{P_{0}}{r^{2}}\left(\langle n_{u}^{(1)}\rangle+\langle n_{u}^{(2)}\rangle+2\langle d_{+}^{(1)}d_{-}^{(2)}\rangle\right)\nonumber \\
 & = & \frac{4P_{0}}{Ar^{2}}\Omega^{2}(4\Delta^{2}+\Omega^{2}+\gamma^{2}).\label{eq:Intensity_forward}
\end{eqnarray}
For convenience in what follows, we present results in terms of a
dimensionless intensity defined by 
\begin{equation}
\mathcal{I\left(\hat{\br}\right)}\equiv\frac{I(\mathbf{r})}{P_{0}/r^{2}}.\label{eq:ScaledIntensity}
\end{equation}
\begin{figure}
\includegraphics[width=0.27\textwidth]{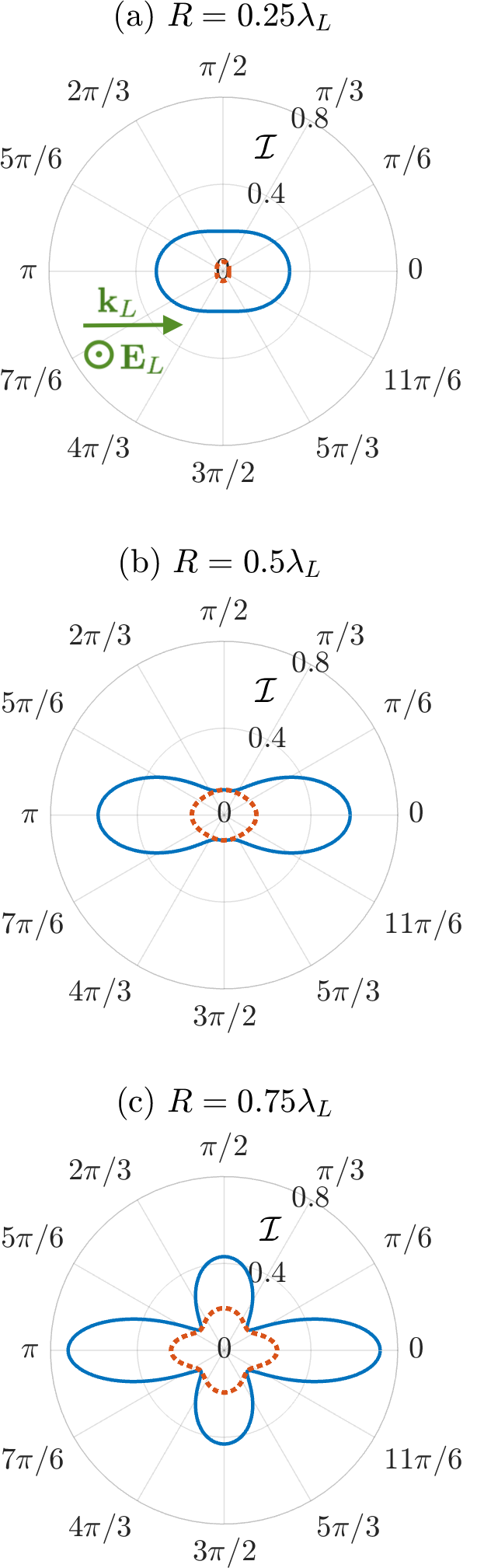}

\caption{\label{fig:PolarPerp}Polar plot of the intensity $\mathcal{I\left(\hat{\br}\right)}$
(solid line) in the $x-y$ plane for (a) $R=0.25\lambda_{L}$; (b)
$R=0.5\lambda_{L}$; (c) $R=0.75\lambda_{L}$. Other parameters are
$\Delta=0,\thinspace\Omega=0.5\gamma$. The dotted line is the incoherent
component of the intensity. \textcolor{black}{The directions of the laser field wavevector and polarisation   are shown in (a) and are the same for all the subfigures.}}
\end{figure}

\begin{figure}
\includegraphics[width=0.48\textwidth]{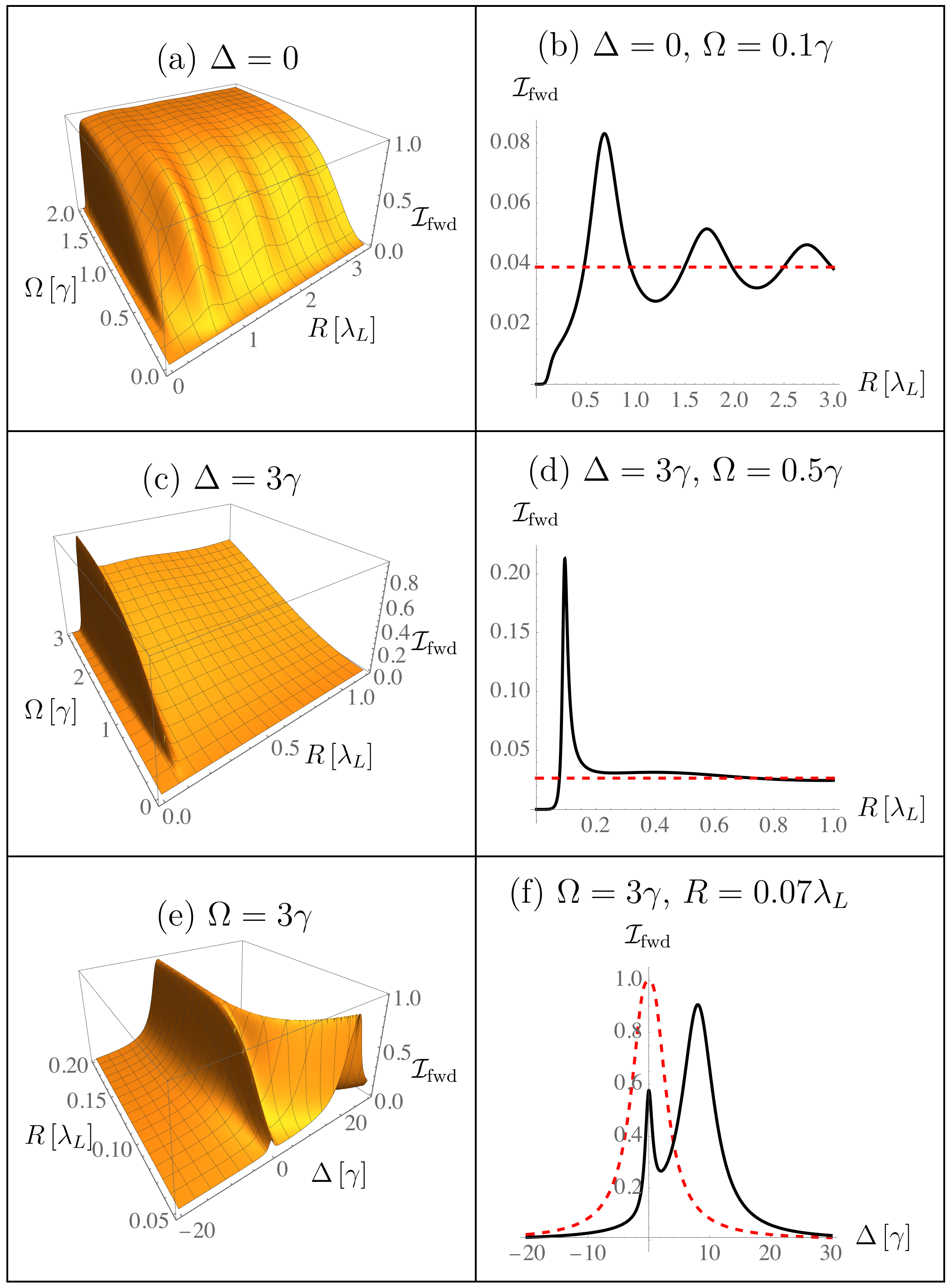}

\caption{\label{fig:Intensity3D_2D}Forward scattered intensity $\mathcal{I_{{\rm fwd}}}$
as a function of $R$, $\Omega$ and $\Delta$, for the perpendicular
configuration. The intensity is scaled as in Eq.(\ref{eq:ScaledIntensity}).
Dashed red line indicates the uncoupled result ${\cal I}_{{\rm fwd}}^{{\rm uc}}$. }
\end{figure}

Examples of the far-field scattered intensity pattern $\mathcal{I\left(\hat{\br}\right)}$
in the $x-y$ plane are shown as polar plots in Fig. \ref{fig:PolarPerp},
where the development of fringes for $R\ge0.5\lambda_{L}$ is clearly
evident, as well as the expected forward-backward symmetry. The incoherent
component of the scattering will be discussed further below. From
Eq.(\ref{eq:I3D_perp_config}) we see that $\mathcal{I\left(\hat{\br}\right)}$
is fully characterised by the quantities $\mathcal{I}{}_{{\rm fwd}}$
and $g_{\perp}^{\left(1\right)}$ where the latter gives the fringe
visibility. The behaviour of $\mathcal{I}{}_{{\rm fwd}}$ is displayed
as a function of $R$, $\Omega$ and $\Delta$ in Fig. \ref{fig:Intensity3D_2D}.
In Fig. \ref{fig:Intensity3D_2D}(a) where $\mathcal{I_{{\rm fwd}}}$
is plotted against $R$ and $\Omega$ for the case $\Delta=0$, two
key features are evident. The first is that for $R\lesssim R_{{\rm nf}}\equiv(k_{L})^{-1}$
(i.e. when the near field terms dominate $G(\mathbf{R})$) the far-field
scattering  \textcolor{black}{decreases below the uncoupled result, tending to zero for small  $R$. We will call this phenomenon the {\em suppression of scattering}. We note though, that for sufficiently large $\Omega$
(outside the range of this plot) appreciable scattering occurs at small $R$.}
The second feature is that for $R>R_{{\rm nf}}$, $\mathcal{I_{{\rm fwd}}}$
oscillates with $R$, but with decreasing amplitude as either $R$
or $\Omega$ increase. In Fig. \ref{fig:Intensity3D_2D}(b), which
is a cross section of Fig. \ref{fig:Intensity3D_2D}(a) at the value
$\Omega=0.1\gamma$, we see that $\mathcal{I_{{\rm fwd}}}$ oscillates
about the uncoupled result $\mathcal{I}_{{\rm fwd}}^{{\rm uc}}$ (see
Eq.(\ref{eq:Ifwd_uc})), with $\mathcal{I_{{\rm fwd}}}\rightarrow\mathcal{I}_{{\rm fwd}}^{{\rm uc}}$
as $R\rightarrow\infty$. Other plots (not shown here) confirm that
in the region satisfying both $\Omega\gtrsim\gamma$ and $R>R_{{\rm nf}}$,
$\mathcal{I_{{\rm fwd}}}\simeq\mathcal{I}_{{\rm fwd}}^{{\rm uc}}$,\textcolor{black}{which
is discussed in more detail below}. Fig. \ref{fig:Intensity3D_2D}(c)
shows $\mathcal{I_{{\rm fwd}}}$ plotted against $R$ and $\Omega$
but now for the case $\Delta=3\gamma$. Once again we observe the
suppression of scattering at small $R$, while throughout the region
$R>R_{{\rm nf}}$ there is little dependence on $R$ with $\mathcal{I_{{\rm fwd}}}\simeq\mathcal{I}_{{\rm fwd}}^{{\rm uc}}$.
The key feature of this graph is the sharp peak of intensity at $R=0.1\lambda_{L}$,
which is very prominently seen in Fig. \ref{fig:Intensity3D_2D}(d)
which is a cross section of Fig. \ref{fig:Intensity3D_2D}(c) at the
value $\Omega=0.5\gamma$. We note, without presenting plots, that
for $\Delta<0$, while suppression of scattering still occurs for
$R\lesssim R_{{\rm nf}},$ there are no sharp peaks such as seen in
Fig. \ref{fig:Intensity3D_2D}(c), and for $R>R_{{\rm nf}}$, $\mathcal{I_{{\rm fwd}}}\simeq\mathcal{I}_{{\rm fwd}}^{{\rm uc}}$.
In Fig. \ref{fig:Intensity3D_2D}(e) we plot $\mathcal{I_{{\rm fwd}}}$
against $\Delta$ and $R$ for the case $\Omega=3\gamma$, and see
that there are two sharp intensity peaks, one which is at $\Delta=0$
for all $R$, and the other which is at $\Delta\simeq24\gamma$ when
$R=0.05\lambda_{L}$ and moves steadily towards $\Delta=0$ as $R$
increases towards $R\simeq R_{{\rm nf}}.$ Fig. \ref{fig:Intensity3D_2D}(f)
which is a cross section of Fig. \ref{fig:Intensity3D_2D}(e) at $R=0.07\lambda_{L}$,
shows that the peak in $\mathcal{I}{}_{{\rm fwd}}$ at $\Delta=0$
is significantly smaller than the (single) peak in $\mathcal{I}{}_{{\rm fwd}}^{{\rm uc}}$.
Eventually, at larger $\Omega$, \textcolor{black}{this} peak in $\mathcal{I}{}_{{\rm fwd}}$
will grow to match the uncoupled result $\mathcal{I}{}_{{\rm fwd}}^{{\rm uc}}$.

The behaviour of $\mathcal{I}{}_{{\rm fwd}}$ discussed above can
be readily understood from its analytic expression (see Eq.(\ref{eq:Intensity_forward}))

\begin{align}
 & \mathcal{I}{}_{{\rm fwd}}=\label{eq:Ifwd_full}\\
 & \frac{4\Omega^{2}(4\Delta^{2}+\Omega^{2}+\gamma^{2})}{\left(\gamma^{2}+4\Delta^{2}\right)\left[(2\text{\ensuremath{G_{i}}(\ensuremath{\mathbf{R}})}+\gamma)^{2}+4(\text{\ensuremath{G_{r}}(\ensuremath{\mathbf{R)}}}+\Delta)^{2}+4\Omega^{2}\right]+4\Omega^{4}}.\nonumber 
\end{align}
\textcolor{black}{First, for large $R$, as noted earlier, $G\text{(\ensuremath{\mathbf{R}})}\rightarrow0$
and $\mathcal{I}{}_{{\rm fwd}}\rightarrow\mathcal{I}{}_{{\rm fwd}}^{{\rm uc}}$.
In fact, for any value of $\Omega$ or $\Delta$, the relative difference
between $\mathcal{I}{}_{{\rm fwd}}$ and $\mathcal{I}{}_{{\rm fwd}}^{{\rm uc}}$
is less than 10\% when $R>5\lambda_{L}$. Furthermore, if either $\Omega\gtrsim\gamma$
or $\Delta\gtrsim\gamma$, the relative difference is less than 10\%
for $R>\lambda_{L}$. The denominator of Eq.(\ref{eq:Ifwd_full})
holds the key to behaviour of $\mathcal{I}{}_{{\rm fwd}}$ with $\Delta$
at small $R$. The factor $\left(\gamma^{2}+4\Delta^{2}\right)$ gives
rise to a resonance at $\Delta=0$ of width $\sim\gamma$, and the
term in square brackets produces a resonance} \footnote{We note that $|2\text{\ensuremath{G_{i}}(\ensuremath{\mathbf{R}})}+\gamma|$
is never smaller than $0.6\gamma$}\textcolor{black}{{} at $\Delta=-\text{\ensuremath{G_{r}}(\ensuremath{\mathbf{R)}}}$
with a power broadened width $\sim\sqrt{\gamma^{2}+\Omega^{2}}$.
We define 
\begin{equation}
\Delta_{d}\equiv-\text{\ensuremath{G_{r}}(\ensuremath{\mathbf{R)}}}\label{eq:diddipshift}
\end{equation}
which is the dipole-dipole shift. In the limit of small $R$ 
\begin{equation}
G(\mathbf{R})\big|_{R\ll\lambda_{L}}\approx\gamma\left(-\frac{3}{4(k_{L}R)^{3}}+\frac{i}{2}\right),\label{eq:GsmallR}
\end{equation}
where the real part of Eq.(\ref{eq:GsmallR}), for which $\mathcal{I}{}_{{\rm fwd}}$
is most sensitive, is very accurate for $R<R_{{\rm nf}}$. Using Eq.(\ref{eq:GsmallR})
in Eq.(\ref{eq:diddipshift}) we find the atomic separation $R_{d}$
which shifts the atoms into resonance with a laser detuning of $\Delta$
is 
\[
R_{d}=\frac{\lambda_{L}}{2\pi}\left(\frac{3\gamma}{4\Delta}\right)^{\frac{1}{3}}
\]
}

\subsection{\textcolor{black}{Suppression of Scattering }}

\textcolor{black}{Suppression of scattering by closely spaced atoms is a well known phenomenon (e.g. \cite{Pellegrino:2014fj}) in which}
\begin{equation}
\eta\equiv\frac{\mathcal{I}{}_{{\rm fwd}}}{\mathcal{I}{}_{{\rm fwd}}^{{\rm uc}}}\ll1.\label{eq:etasubrad}
\end{equation}
\textcolor{black}{We will define the regime of suppression of scattering to be where  $\eta\lesssim0.1$,  and inserting Eq.(\ref{eq:Ifwd_full}) and
the dimensionless form of Eq.(\ref{eq:Ifwd_uc}) into Eq.(\ref{eq:etasubrad}) we find this criteria is satisfied when }
\begin{equation}
\text{\ensuremath{|G_{r}}(\ensuremath{\mathbf{R)}}}+\Delta|\geq\sqrt{\frac{\gamma^{2}}{4}+\Delta^{2}+\frac{\Omega^{4}}{\left(\gamma^{2}+4\Delta^{2}\right)}}\;.\label{eq:subradcond}
\end{equation}
\textcolor{black}{Since suppression of radiation occurs
only at small $R$, we can use
the approximation $G_{r}(\mathbf{R})\approx-3\gamma/4(k_{L}R)^{3}$
in Eq.(\ref{eq:subradcond})}. 

\textcolor{black}{In the regime of suppressed radiation we find } 
\begin{equation}
\mathcal{I}_{\rm fwd}|_{\Delta=0}\simeq\frac{\Omega^{2}(\Omega^{2}+\gamma^{2})}{\gamma^{2}\Delta_{d}^{2}+\Omega^{4}},\label{eq:Ifwd_Del0}
\end{equation}
and
\begin{equation}
\eta|_{\Delta=0}\simeq\frac{\left(2\Omega{}^{2}+\gamma^{2}\right)^{2}}{4\left(\gamma^{2}\Delta_{d}^{2}+\Omega^{4}\right)}.\label{eq:Ifwd_uc_Del0}
\end{equation}
It is also interesting to note that when $|\Delta_{d}|\gg\gamma$
and $\gamma\gg\Omega$ 
\begin{equation}
\mathcal{I}{}_{{\rm fwd}}|_{\Delta=0}\simeq\Omega^{2}/\Delta_{d}^{2}\label{eq:swapresonance}
\end{equation}
while 
\begin{equation}
\mathcal{I}{}_{{\rm fwd}}|_{\Delta=\Delta_{d}}\simeq\frac{\Omega^{2}}{\gamma^{2}+\Omega^{2}}.\label{eq:Ifwed_DeleqDeld}
\end{equation}

\subsection{Incoherent scattering}

\begin{figure}
\textcolor{black}{\includegraphics[width=0.48\textwidth]{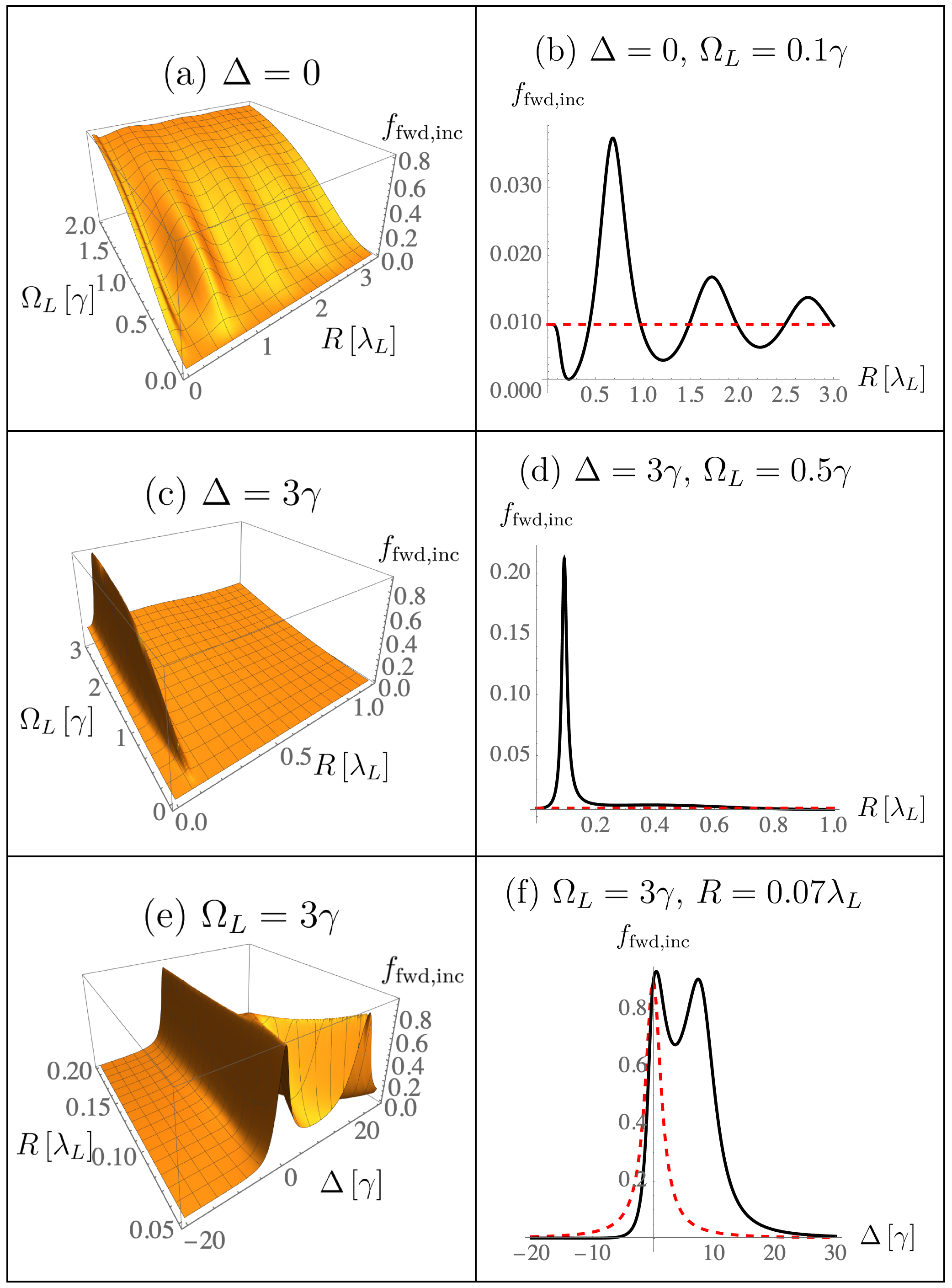}}

\textcolor{black}{\caption{\textcolor{black}{\label{fig:Intensityinc3D_2D}Incoherent fraction
of forward scattered intensity as a function of $R$, $\Omega$ and
$\Delta$, for the perpendicular configuration. Dashed red line is
$f_{{\rm fwd,inc}}^{{\rm uc}}$. }}
}
\end{figure}

In Fig. \ref{fig:PolarPerp} the incoherent component of the scattered
intensity displays spatial interference fringes, and their origin
can be readily determined from the expression for the incoherent scattered
intensity

\begin{align}
\mathcal{I}_{\text{\text{inc}}} & (\hat{\mathbf{r}})=2\left(1-\frac{r_{z}^{2}}{r^{2}}\right)\left[\langle n_{u}^{(1)}\rangle-\langle d_{+}^{(1)}\rangle\langle d_{-}^{(1}\rangle+\Re\left(e^{-ik_{L}\hat{\mathbf{r}}\cdot\mathbf{R}}\chi_{d}\right)\right].\label{eq:Iinc_def}
\end{align}
In Eq.(\ref{eq:Iinc_def}), the spatial dependence in the horizontal
plane arises from $\chi_{d}$, the incoherent component of the dipole
correlation, which has the relatively simple form
\begin{align}
 & \chi_{d}\equiv\langle d_{+}^{(1)}d_{-}^{(2)}\rangle-\langle d_{+}^{(1)}\rangle\langle d_{-}^{(2)}\rangle\\
 & =-\frac{8\Omega^{4}\left(\gamma^{2}+4\Delta^{2}\right)\left(\ensuremath{G_{i}}(\ensuremath{\mathbf{R}})\gamma+2\ensuremath{G_{r}}(\ensuremath{\mathbf{R)}}\Delta\right)}{A^{2}}.\label{eq:Xdsol}
\end{align}
Eq.(\ref{eq:Xdsol}) shows clearly that it is the dipole coupling
that causes $\chi_{d}$ to be non zero. The visibility of the incoherent
fringes is directly dependent on $\chi_{d}$, and is given by
\begin{align}
V_{{\rm inc}} & =\frac{\left|\chi_{d}\right|}{\langle n_{u}^{(1)}\rangle-\langle d_{+}^{(1)}\rangle\langle d_{-}^{(1)}\rangle}\nonumber \\
 & =\frac{4\left|G_{i}(\ensuremath{\mathbf{R}})\gamma+2G_{r}(\ensuremath{\mathbf{R}})\Delta\right|(\gamma^{2}+4\Delta^{2})}{4|G(\ensuremath{\mathbf{R}})|^{2}(\gamma^{2}+4\Delta^{2})+(\gamma^{2}+4\Delta^{2}+2\Omega^{2})^{2}}.\label{eq:Vinc}
\end{align}
From Eq.(\ref{eq:Vinc}) we find that incoherent fringes are visible
only in a narrow regime, with $\Omega\lesssim\gamma$, $\Delta\lesssim\gamma$,
and $|G(\mathbf{R})|\approx\gamma$ (i.e. $R\approx\lambda_{L}$). The behaviour
of the incoherent fraction of the forward scattering, $f_{\text{fwd,inc}}$,
is displayed in Fig. \ref{fig:Intensityinc3D_2D} for the same parameters
as in Fig. \ref{fig:Intensity3D_2D}, and we see that $f_{\text{fwd,inc}}$
broadly follows $\mathcal{I}_{\text{fwd}}$, apart from the region
$R\rightarrow0$. Here at $\Delta=0$ in the regime of \textcolor{black}{suppressed scattering} (see
Eq.(\ref{eq:subradcond})), $\mathcal{I}_{\text{fwd}}\rightarrow0$
while $f_{\text{fwd,inc}}\rightarrow f_{{\rm fwd,inc}}^{{\rm uc}}$,
as shown in \ref{fig:Intensityinc3D_2D} (a),(b), and in \ref{fig:Intensityinc3D_2D}
(e),(f). As expected, we also find $f_{\text{fwd,inc}}\rightarrow f_{{\rm fwd,inc}}^{{\rm uc}}$
in the regime of large $R$.

\textcolor{black}{}

\section{Decorrelation Approximation\label{subsec:Decorrelation-Approximation}}

The analytic solution given in Section \ref{sec:Formalism} provides
a comprehensive description of the steady-state behaviour of the driven
two atom system over a full range of $\Omega,\Delta$ and $R$, \textcolor{black}{which we have explored in detail in Section \ref{sec:Results_Perp}.}  It
is a formidable challenge however to obtain a comparably detailed
description for a larger atomic ensemble because the number of required
equations scales as $\sim2^{2N}$, where $N$ is the number of atoms.
In this section we present an approximate solution method for a system
of dipole coupled atoms, which has a more favorable scaling with atom
number, and has the potential for solving the behaviour of larger
systems. \textcolor{black}{By comparing the results of the approximate solution of  Eqs.(\ref{eq:dlu}) - (\ref{eq:nunu}) to our exact analytic solution, we are able to characterise  in detail the validity and accuracy of the approximate solution. } 

 \textcolor{black}{The decorrelation approximation we present below is  familiar in the field of quantum optics, and others, and can be viewed as a low order truncation of a cumulant expansion.  A recent example has been given by  Kr\"{a}mer and Ritsch \cite{Kramer:2015fm}, who have numerically analysed the effects of first and second order decorrelations on the temporal evolution of an array of vacuum coupled, undriven, two-state atoms. We show that for the case of two driven atoms,  the approximation is accurate over a wide range of parameters, including the strongly driven regime. We obtain an analytic description of the validity regime, and also use our methodology to  introduce the concept of an  \emph{effective driving field}, which we shall see provides additional physical insight into the behaviour of the two atom system.}

\subsection{The Decorrelated Equations }

We begin by defining an effective driving field for atom $i$
which is the sum of two fields arriving at atom $i$, namely the laser
field and the field scattered from the other atom $j$. For convenience
we will express this in terms of an effective Rabi frequency

\begin{equation}
\Omega_{\text{Eff}}^{(i)}=e^{i\mathbf{k}_{L}\cdot{\bf r}_{i}}\Omega+2G(\mathbf{R})d_{-}^{(j)};\ \ (j\ne i)\label{eq:Eeff_def}
\end{equation}
 and we note that we have dropped the quantum noise term (e.g. see
Eq.(\ref{eq:E_a_free}) ) since it disappears in all expectation values
that we take. Eqs.(\ref{eq:dlu}) and (\ref{eq:nu}) then take the
form 

\begin{align}
-i\frac{d}{dt}\langle d_{-}^{(i)}\rangle= & \left(\Delta+i\frac{\gamma}{2}\right)\langle d_{-}^{(i)}\rangle+\frac{1}{2}\langle\Omega_{\text{Eff}}^{(i)}(1-2n_{u}^{(i)})\rangle,\label{eq:dlu_Eeff}
\end{align}

and

\begin{align}
-i\frac{d}{dt}\langle n_{u}^{(i)}\rangle= & i\gamma\langle n_{u}^{(i)}\rangle+\frac{1}{2}\left(\langle\Omega_{\text{Eff}}^{(i)}d_{+}^{(i)}\rangle-\langle\left(\Omega_{\text{Eff}}^{(i)}\right)^{\dagger}d_{-}^{(i)}\rangle\right).\label{eq:nu_Eeff}
\end{align}

We now make the decorrelation approximation
\begin{equation}
\langle d_{\pm}^{\left(i\right)}n_{u}^{\left(j\right)}\rangle\rightarrow\langle d_{\pm}^{\left(i\right)}\rangle\langle n_{u}^{\left(j\right)}\rangle\label{eq:Decorr1}
\end{equation}
\begin{equation}
\langle d_{\pm}^{\left(i\right)}d_{\mp}^{\left(j\right)}\rangle\rightarrow\langle d_{\pm}^{\left(i\right)}\rangle\langle d_{\mp}^{\left(j\right)}\rangle\label{eq:Decorr2}
\end{equation}
(for $i\ne j$) which decouples Eqs.(\ref{eq:dlu}) (and its conjugate)
and (\ref{eq:nu}) from Eqs.(\ref{eq:dludul})-(\ref{eq:nunu}) leaving
us with a set of three approximate equations for the system:

\begin{align}
-i\frac{d}{dt}\langle d_{-}^{(i)}\rangle_{D}= & \left(\Delta+i\frac{\gamma}{2}\right)\langle d_{-}^{(i)}\rangle_{D}-\frac{\langle\Omega_{\text{Eff}}^{(i)}\rangle_{D}}{2}\left(1-2\langle n_{u}^{(i)}\rangle_{D}\right),\label{eq:dlu_Eeffdec}
\end{align}

\begin{align}
-i\frac{d}{dt}\langle n_{u}^{(i)}\rangle_{D}= & i\gamma\langle n_{u}^{(i)}\rangle_{D}\label{eq:nu_Eeffdec}\\
 & +\frac{\langle\Omega_{\text{Eff}}^{(i)}\rangle_{D}}{2}\langle d_{+}^{(i)}\rangle_{D}-\frac{\langle\left(\Omega_{\text{Eff}}^{(i)}\right)^{\dagger}\rangle_{D}}{2}\langle d_{-}^{(i)}\rangle_{D},\nonumber 
\end{align}
and the conjugate of Eq.(\ref{eq:dlu_Eeffdec}). In these equations
the subscript $D$ indicates that these are the decorrelated expectation
values. We note that while the number of system equations has been
reduced (scaling as $\sim N$ for $N$atoms), they are now nonlinear
due to the effective field's dependence on $\langle d_{-}^{j}\rangle_{D}$.
Eqs.(\ref{eq:dlu_Eeffdec}) and (\ref{eq:nu_Eeffdec}) have the same
form as the original Mollow equations\cite{Mollow:1969wc} for a
single driven atom, but with the substitution $\Omega\rightarrow\langle\Omega_{\text{Eff}}^{(1)}\rangle_{D}$.
The solutions to Eqs.(\ref{eq:dlu_Eeffdec}) and (\ref{eq:nu_Eeffdec})
can be written formally using the expressions in Eqs.(\ref{eq:nu_Mollow})
and (\ref{eq:dlu_Mollow}), although this is of formal rather than
practical value, as $\langle\Omega_{\text{Eff}}^{(1)}\rangle_{D}$
must of course be found as part of the solution, which here is carried
out numerically.

\begin{figure}
\centering{}

\includegraphics[scale=0.35]{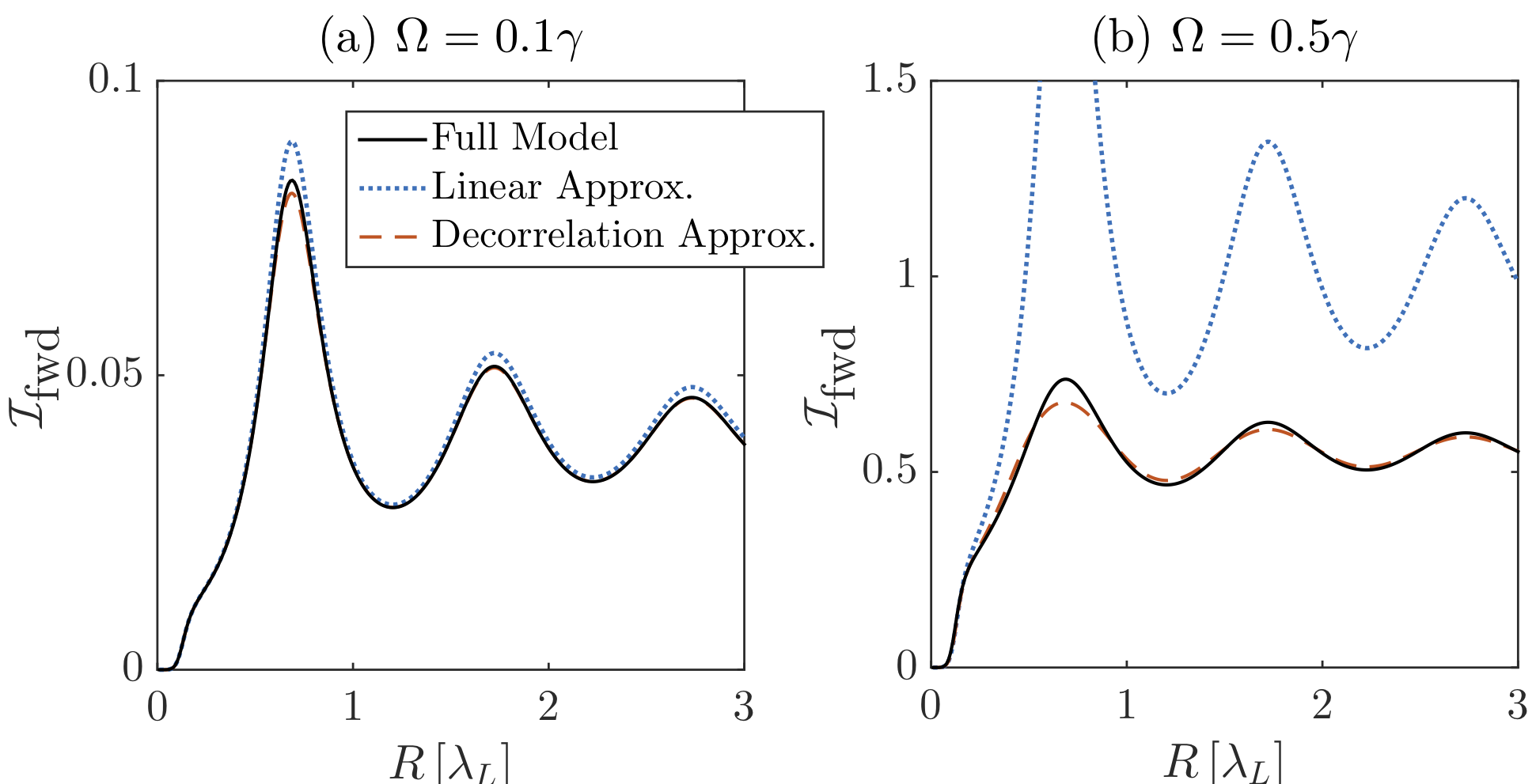}

$\qquad$

\includegraphics[scale=0.35]{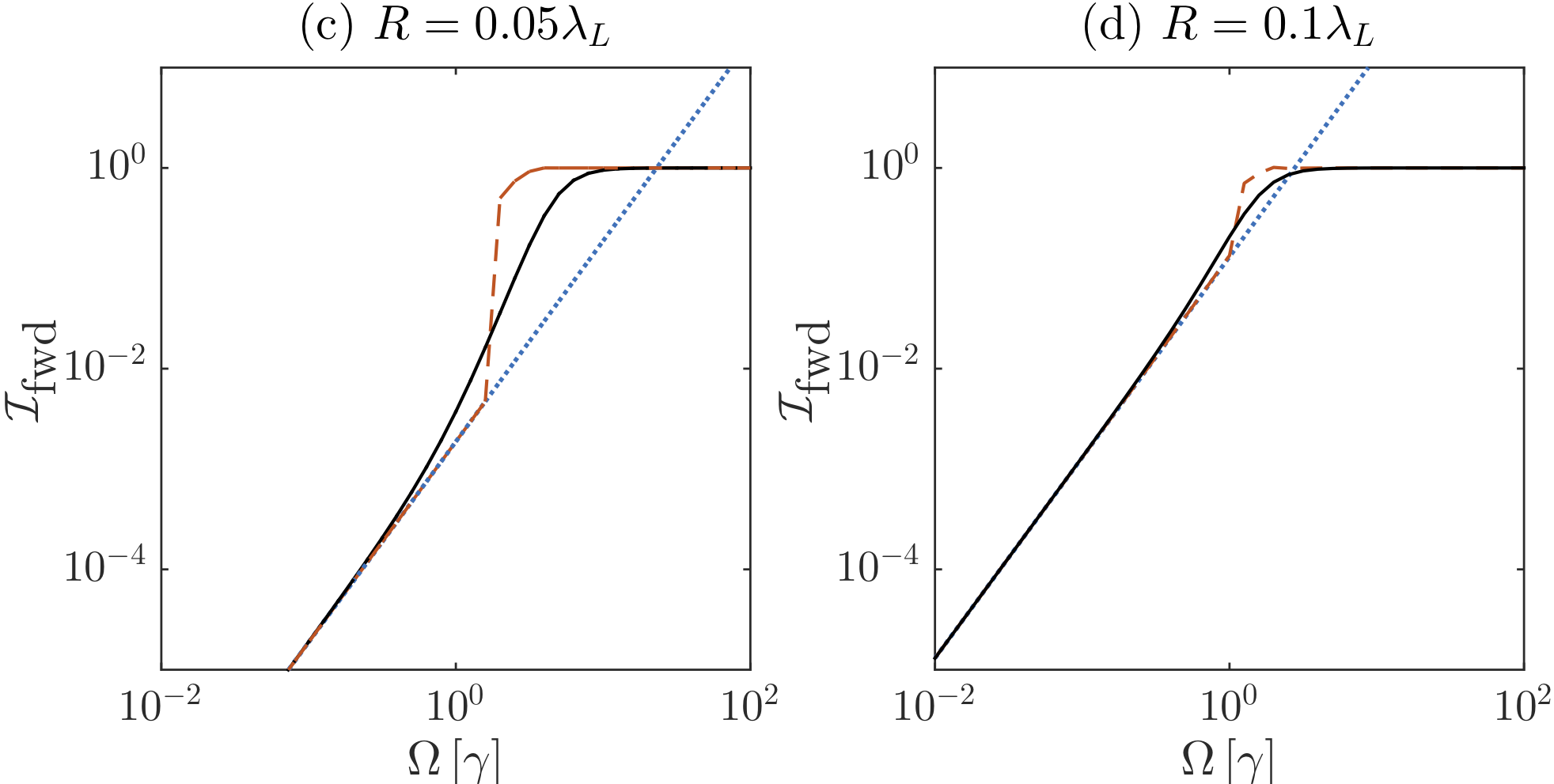}

\caption{\label{fig:Compare_withR}Comparison of ${\cal I}_{{\rm fwd,}D}$
(dashed red line) and ${\cal I}_{{\rm fwd}}$ (solid line) for the
case of $\Delta=0$. The dotted line is the solution for the forward
intensity obtained using the linear approximation.}
\end{figure}

In Fig. \ref{fig:Compare_withR} we compare the forward scattered
intensity obtained from the decorrelated equations, ${\cal I}_{{\rm fwd,}D}$,
with the true forward scattered intensity, for the case of $\Delta=0$.
We also include a comparison to the commonly employed linear approximation
(e.g. \cite{Zhu:2016ds,Pellegrino:2014fj}), which is obtained by
setting $n_{u}^{(i)}=0$ in Eqs.(\ref{eq:dlu})-(\ref{eq:nunu}). 

We see from Figs. \ref{fig:Compare_withR}(a) and (b) that at $\Delta=0$,
the decorrelation approximation provides a good solution for all $R$
for the values of $\Omega$ shown. We also see that the linear approximation
is accurate at $\Omega=0.1\gamma$ but fails badly by $\Omega=0.5\gamma$.
Figs. \ref{fig:Compare_withR}(c) and (d), where $\mathcal{I}{}_{{\rm fwd}}$
is plotted against $\Omega$ in the challenging regime of small $R$,
show that the decorrelation solution provides an accurate representation
in both the low and high intensity regimes, but is less accurate in
the transition region around $\Omega=\gamma.$ Once again, as expected,
the linear approximation is shown to be poor for $\Omega\apprge0.5\gamma$.
We will obtain a quantitative expression for the validity regime in
the next section.

\subsection{\textcolor{black}{Validity regime\label{subsec:Validity-regime}}}

\begin{figure}
\centering{}\includegraphics[scale=0.43]{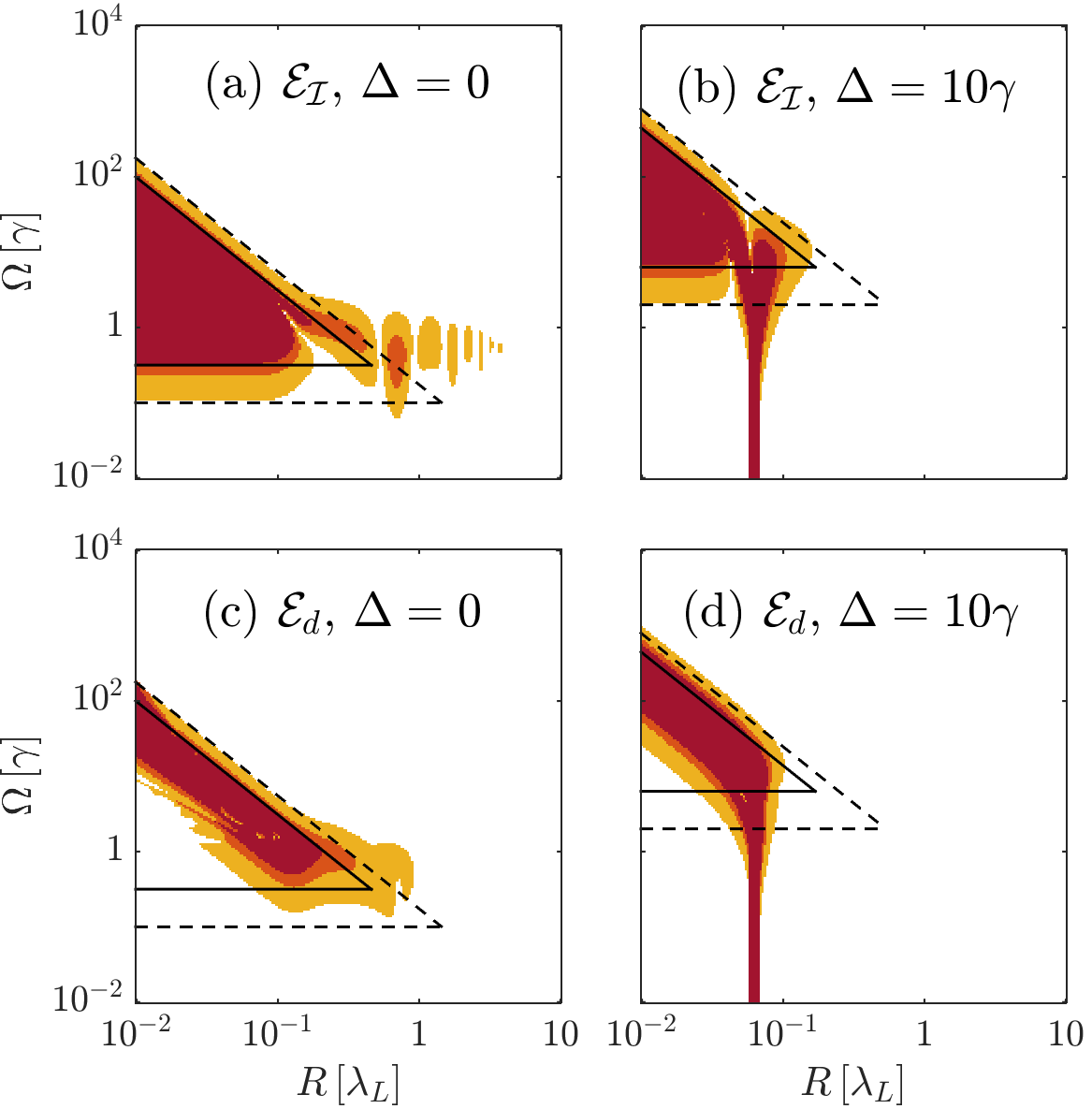}\caption{\label{fig:error_with_inc}Relative errors arising from the decorrelation
approximation. (a), (b) $\mathcal{E_{I}}$ ; (c),(d) $\mathcal{E}_{d}$
. Yellow regions, $0.01<{\cal E}<0.05$. Orange regions, $0.05<{\cal E}<0.1$.
Red regions, $0.1<{\cal E}$. The black solid (dashed) lines give
the boundary of the ${\cal E}>0.1$ ($0.01$) regions as approximated
by Eqs. (\ref{eq:Horizbdry},\ref{eq:approx_validregime-1}).}
\end{figure}

The relative error in the forward scattered intensity due to the decorrelation
approximation is given by 

\begin{align}
{\cal E}_{{\cal I}} & \equiv\left|\frac{{\cal I}_{{\rm fwd}}-{\cal I}_{{\rm fwd,}D}}{{\cal I}_{{\rm fwd}}}\right|.\label{eq:Er}
\end{align}
The quantity ${\cal E}_{{\cal I}}$ is plotted for a wide range of
$R$ and $\Omega$ in the top row of Fig. \ref{fig:error_with_inc}
for representative detunings (a) $\Delta=0$ and (b) $\Delta=10$.
In this plot, we see that the region of significant error (e.g. ${\cal E}_{{\cal I}}>0.01)$
is essentially confined to a triangular area in $\log R$ and $\log\Omega$,
with the particulars of the area dependent on the value of $\Delta$.
For $\Delta>0$, an additional thin `tail' emerges in the low $\Omega$
region at the value $R=R_{d}$, as seen in \ref{fig:error_with_inc}(b).
In this tail region the atoms have been pulled into resonance with
the driving field, and the effective field is large. The relative
error of the mean dipole due to the decorrelation approximation is
given by
\begin{equation}
{\cal E}_{d}\equiv\left|\frac{\langle d_{-}^{(1)}\rangle-\langle d_{-}^{(1)}\rangle_{D}}{\langle d_{-}^{(1)}\rangle}\right|\label{eq:Errordipole}
\end{equation}
 and this quantity is plotted in the second row of Fig. \ref{fig:error_with_inc}.
We see that ${\cal E}_{d}$ is significant only within the same region
as ${\cal E}_{{\cal I}}$. Corresponding plots (not shown here) for
the relative error of $\langle n_{u}^{(1)}\rangle_{D}$ or $\langle\Omega_{\text{Eff}}^{(1)}\rangle_{D}$
are very similar to the plot of ${\cal E}_{{\cal I}}$. Using this
fact we can find a useful analytic expression for the validity range
of the decorrelation approximation. Noting first that 
\begin{equation}
{\cal I}_{{\rm fwd,}D}\left(\Omega\right)={\cal I}_{{\rm fwd}}^{{\rm uc}}\left(\langle\Omega_{\text{Eff}}^{(1)}\rangle_{D}\right)\label{eq:IfwdDIfwduc}
\end{equation}
(see discussion following Eq.(\ref{eq:nu_Eeffdec})), we make the
substitution $\langle\Omega_{\text{Eff}}^{(1)}\rangle_{D}\rightarrow\langle\Omega_{\text{Eff}}^{(1)}\rangle$
in the RHS of Eq.(\ref{eq:IfwdDIfwduc}), then evaluate $\langle\Omega_{\text{Eff}}^{(1)}\rangle$
using Eq.(\ref{eq:dlu_analytic}), to give an analytic expression
for ${\cal E}_{{\cal I}}$ that is valid in the region where ${\cal E}_{{\cal I}}$
is small. Further simplification can be obtained by replacing $G(\mathbf{R})$
by its approximate form Eq.(\ref{eq:GsmallR}), which is valid wherever
${\cal E}_{{\cal I}}$ is significant. \textcolor{black}{We find the
lower boundary of the validity range by noting that in this region,
when $R<R_{{\rm nf}}$, then $\left(kR\right)^{-3}\gg\left(\Omega/\gamma\right)^{2}\thinspace,\thinspace\left(\Delta/\gamma\right)^{2}$
so we can take $\lim_{R\rightarrow0}{\cal E}_{{\cal I}}$ to obtain
\begin{equation}
{\cal E}_{{\cal I}}\approx\frac{\Omega^{2}}{\left(\gamma^{2}+4\Delta^{2}\right)}\Theta\label{eq:Horizbdry}
\end{equation}
where
\begin{equation}
\Theta=\frac{\left|3\gamma^{2}+4(\Omega^{2}-5\Delta^{2})\right|}{\gamma^{2}+4\Delta^{2}+\Omega^{2}}.
\end{equation}
Eq.(\ref{eq:Horizbdry}) describes horizontal lines in the $\left(\log R,\log\Omega\right)$
plane, of defined value of relative error. The quantity $\Theta$
is bounded by the value 5, but for most of parameter space it is much
less, and we find that setting $\Theta=1$ gives good agreement with
our numerical results. Two example contours ($\Omega^{2}/\left(\gamma^{2}+4\Delta^{2}\right)=0.01;\Omega^{2}/\left(\gamma^{2}+4\Delta^{2}\right)=0.1)$
are plotted on the subfigures in Fig. \ref{fig:error_with_inc} and
are seen to accurately capture the small $\Omega$ boundary of the
validity range of the decorrelation approximation out to $R\simeq R_{{\rm nf}}$,
apart from the resonance tail. At the upper boundary, in the region
$R<R_{{\rm nf}}$, we have $|G(\mathbf{R})|,\Omega\gg\gamma,|\Delta|$
. Assuming in addition that $\Omega^{3}(k_{L}R)^{3}\gg\left(1+4\Delta^{2}/\gamma^{2}\right)$
we find
\begin{equation}
{\cal E}_{{\cal I}}\approx\frac{9}{16\Omega^{4}}\frac{\gamma^{2}(\gamma^{2}+4\Delta^{2})}{(k_{L}R)^{6}}.\label{eq:approx_validregime-1}
\end{equation}
Eq.(\ref{eq:approx_validregime-1}) describes diagonal lines in the
$\left(\log R,\log\Omega\right)$ plane, of defined value of relative
error. Two example contours from this equation are shown in Fig. \ref{fig:error_with_inc}
and are seen to give excellent agreement with the large $\Omega$
boundary of the decorrelation approximation out to $R\simeq R_{{\rm nf}}$.
While Eqs.(\ref{eq:Horizbdry}) and (\ref{eq:approx_validregime-1})
are only strictly valid in the range $R<R_{{\rm nf}}$, we have extended
the lines to where they meet at $R\approx\lambda_{L}$ outlining a
triangle which defines a practical validity regime for the decorrelation
approximation, apart from the resonance tail, which occurs at $R=R_{d}$. }

\subsection{\textcolor{black}{Role of the effective field \label{subsec:Effective-field-interpretation}}}

\begin{figure}
\centering{}\centering{}\textcolor{black}{\includegraphics[width=0.45\textwidth]{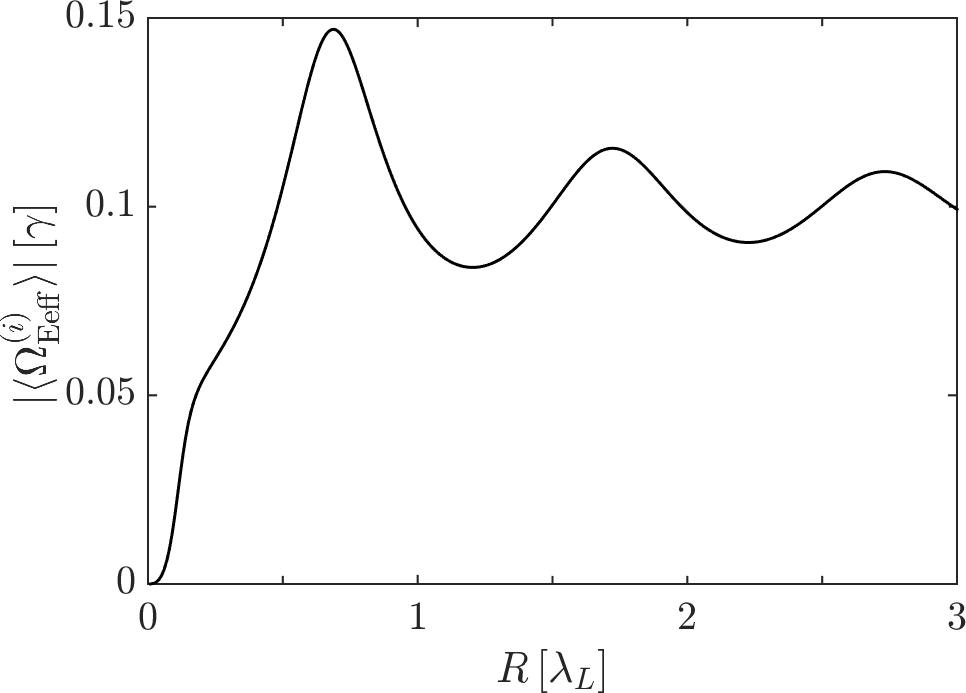}\caption{\label{fig:Perp_Eeff}Effective field for two atoms in the perpendicular
configuration, as a function of interatomic distance $R$. Parameters
same as Fig.\ref{fig:Intensity3D_2D}(b). }
} 
\end{figure}

The mean value of the effective field is the coherent part of the
total field driving each atom, and in the region where the decorrelation
approximation is valid, ${\cal I}_{{\rm fwd}}$ can be obtained by
replacing the laser field in ${\cal I}_{{\rm fwd}}^{{\rm uc}}$ with
the effective field. In Fig.\ref{fig:Perp_Eeff} the magnitude of
the effective field is plotted against $R$ for the same parameters
as Fig.\ref{fig:Intensity3D_2D}(b), and two key features emerge:
(i) the magnitude of the effective field goes to $0$ for small $R$;
(ii) the effective field oscillates with $R$ for $R\gtrsim\lambda_{L}/2$.
Comparison with Fig.\ref{fig:Intensity3D_2D}(b) illustrates that
the scattered field ${\cal I_{{\rm fwd}}}$ broadly follows the magnitude
of the effective field.

\textcolor{black}{The oscillation in the effective field with $R$
is due to interference between its two constituents, the laser field
and the scattered field. Eq.(\ref{eq:Eeff_def}) shows that at atom
1 the phase of the scattered field relative to the laser field arises
from $\langle d_{-}^{\left(2\right)}\rangle$ and $G(\mathbf{R})$.
In the regime $R\gtrsim\lambda_{L}/2$ the phase of $G(\mathbf{R})$
is primarily determined by the factor $e^{ik_{L}R}$ (see Eq.(\ref{eq:arbgdef})).
The phase of $\langle d_{-}^{\left(2\right)}\rangle$ consists of
the phase of the laser field at atom 2, plus an additional shift,
which on resonance and in the regime $R\gtrsim\lambda_{L}/2$ is near
$\pi/2$ (see Eq.(\ref{eq:dlu_analytic})). Thus in the perpendicular
configuration the net phase difference between the two constituent
fields is close to $k_{L}R+\pi/2$, leading to a modulation of the
effective field seen in }Fig.\ref{fig:Perp_Eeff}\textcolor{black}{,
which has a period of approximately $\lambda_{L}$.} \textcolor{black}{The
effective field concept also allows us to understand the \textcolor{black}{suppression of  scattering
 as $R\rightarrow0$. Here, the near complete
destructive interference between the laser field and the scattered
field from the other atom reduces the  effective driving field to near zero }. An analytic
expression for the behaviour of the effective field at small $R$
can be obtained by evaluating Eq.(\ref{eq:Eeff_def})  in the \textcolor{black}{suppressed scattering}
regime to give
\begin{equation}
\langle\Omega_{\text{Eff}}^{(1,2)}\rangle{}_{R\rightarrow0}\approx\frac{2\Omega(2\Omega+(\gamma-2i\Delta)^{2})}{3\gamma^{2}(\gamma-2i\Delta)}(k_{L}R)^{3}.
\end{equation}
}

\section{Results: Parallel Configuration\label{sec:Other-Geometries}}

\begin{figure}
\begin{centering}
\includegraphics[width=0.35\textwidth]{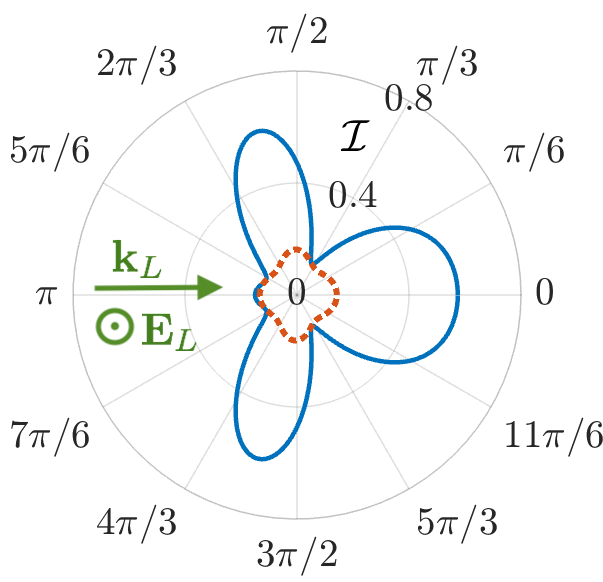}\caption{\label{fig:PolarPara}Polar plot of the intensity $\mathcal{I\left(\hat{\br}\right)}$
(solid line) in the $x-y$ plane scattered from two atoms in the parallel configuration.
The dotted line is the incoherent component of the intensity. \textcolor{black}{ The directions of the laser field wavevector and polarisation  are indicated in green}. Parameters are $R=0.75\lambda_{L}$, $\Delta=0$ and $\Omega=0.5\gamma$.}
\par\end{centering}
\end{figure}

\begin{figure}[H]
\[
\qquad
\]

\begin{centering}
\includegraphics[width=0.45\textwidth]{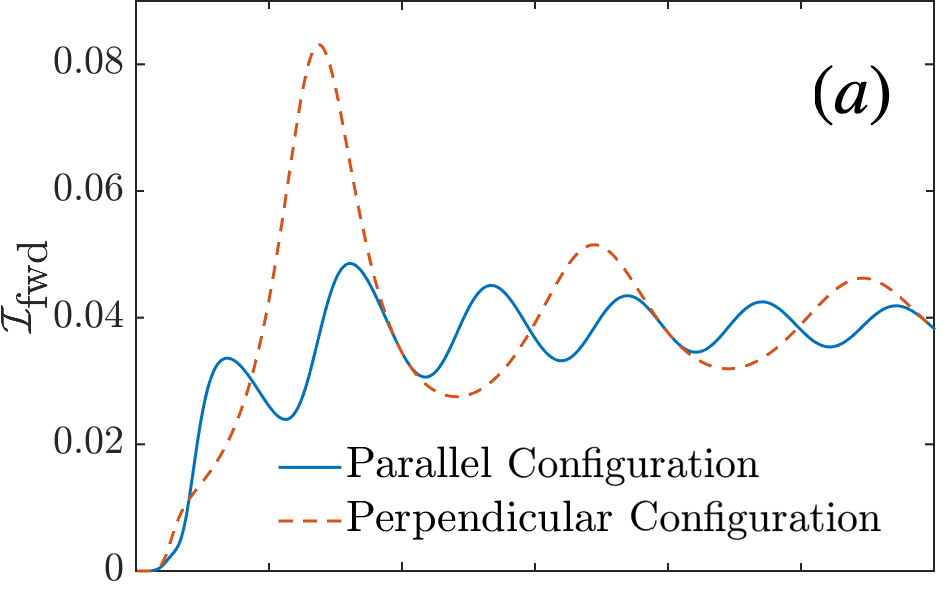}
\par\end{centering}
\centering{}\includegraphics[width=0.45\textwidth]{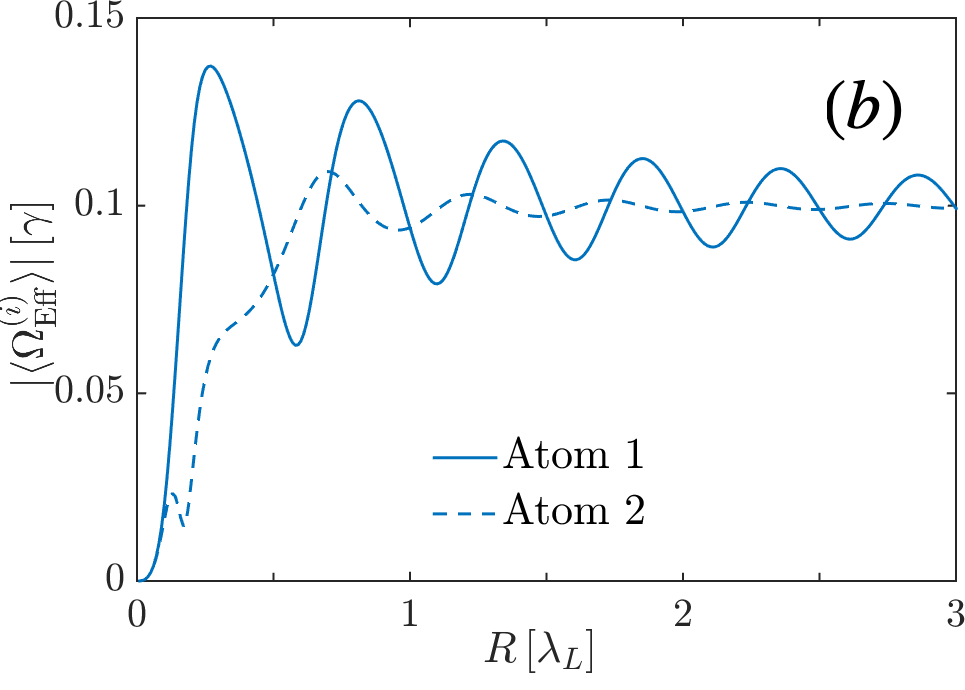}\caption{\label{fig:ParAxis_Imax}(a) Comparison of ${\cal I}_{{\rm fwd}}$
solutions for perpendicular and parallel configurations. Parameters
are $\Omega=0.1\gamma$, $\Delta=0.$ (b) Effective field magnitude
as a function of $R$ for the parallel configuration.  Parameters
same as (a).}
\end{figure}

Finally we consider the case where $\mathbf{R}\parallel\mathbf{k}_{L}$,
with $\mathbf{r}_{1}=-\frac{R}{2}\hat{\mathbf{x}}$ and $\mathbf{r}_{2}=\frac{R}{2}\hat{\mathbf{x}}$
which we call the\textit{\textcolor{black}{{} parallel configuration}} \textcolor{black}{(see Fig. \ref{fig:System})} .
As before we assume $\mathbf{E}_{L}\parallel\hat{\mathbf{z}}$ so
that only two states of each atom participate, but now the atoms are
no longer symmetric with respect to the laser, and for example $\langle d_{-}^{(1)}\rangle\ne\langle d_{-}^{(2)}\rangle$.
A numerically calculated example of the scattered far-field intensity
pattern $\mathcal{I\left(\hat{\br}\right)}$ in the $x-y$ plane is
shown as a polar plot in Fig. \ref{fig:PolarPara}, for the same parameters
as in Fig. \ref{fig:PolarPerp}(c). The salient difference between
these two Figures is that in the parallel configuration the scattering
has developed a forward asymmetry due to phase matching. (For $R=n\lambda_{l}/2,\thinspace n=1,2,...\thinspace,$ the
scattering retains forward-backward symmetry). The forward scattered
intensity ${\cal I}_{{\rm fwd}}$ for the case in Fig. \ref{fig:Intensity3D_2D}(b)
is plotted against $R$ in Fig. \ref{fig:ParAxis_Imax}(a), along
with the corresponding result from the perpendicular configuration.
Both configurations show subradiant scattering as $R\rightarrow0$,
but the oscillations in ${\cal I}_{{\rm fwd}}$ for the parallel configuration
are approximately half the magnitude and twice the frequency of the
perpendicular configuration. This difference is explained by the asymmetry
between the effective fields of the two atoms in the parallel configuration,
which is evident in Fig. \ref{fig:ParAxis_Imax}(b). The scattered
field incident on atom 1 from atom 2 has phase $k_{L}R$ relative
to dipole 2, which itself has a phase $k_{L}R+\pi/2$ relative to
the laser field at atom 1. The net phase difference of $2k_{L}R+\pi/2$
leads to a modulation period of $\lambda_{L}/2$ for the effective
field at atom 1. However at atom 2, the scattered field and the laser
field have travelled the same additional path length from atom 1,
so the phase difference is mainly due to the $\pi/2$ advance from
the resonantly driven atom 1. This means that for $R\gtrsim\lambda_{L}/2$
the effective field driving atom 2 is only weakly modulated as $R$
changes, and the oscillations in $\mathcal{I}_{\text{fwd}}$ shown
in Fig. \ref{fig:ParAxis_Imax}(a) are half the amplitude of those
for the perpendicular configuration, because only one atom contributes
to them. 

\section{Conclusion}

We have used a rigorous formalism to describe the collective scattering
behaviour of two stationary $j_{l}=0\leftrightarrow j_{u}=1$ atoms
driven by a monochromatic laser, and interacting via the vacuum field.
With a suitable choice of system geometry and laser polarisation,
the atoms reduce to effectively two-state. A further restriction to
the perpendicular configuration enabled an analytic solution to be
found for the steady-state mean atomic values and second order correlations
for this system, and hence for the spatial behaviour of the steady-state
far-field scattered intensity. These analytic solutions are valid
over a full range of the parameters $\Omega,\Delta$ and $R$, and
have facilitated a unified and comprehensive survey of the steady-state
scattering behaviour. Key features have been identified and quantified
, including the two resonance peaks for the forward scattered intensity
at $\Delta=0$ and $\Delta=\Delta_{d}$. The incoherent component
of the scattering was shown to exhibit spatial fringes in a defined
regime, underscoring the fact that at small atomic separation, spontaneous
photons are emitted from the joint two-atom system, rather than independently
from each atom. By comparing the expression for the collective scattered
intensity with the corresponding intensity from two uncoupled atoms,
a precise specification has been given of the regime where the dipole-dipole
coupling has significant effect. In particular a simple analytic expression
for the regime of \textcolor{black}{suppressed scattering} and the magnitude of  \textcolor{black}{that} suppression has been derived. 

It is unlikely that a useful analytic solution can be obtained for
the case of more than two atoms, because of the unfavorable scaling
of the number of system equations with the number of atoms. Therefore,
with the aim of finding a solution method with potential application to a large
number of atoms, we have explored an approximation scheme which
has much more favorable scaling. We have shown, with a detailed analysis
on the two-atom system, that our decorrelation approximation provides
an accurate solution for a wide range of the parameters $\Omega,\Delta$
and $R$, and we have given an analytic description of the validity
regime. Finally, the concept of the effective driving field has been
shown to provide a direct physical interpretation of key aspects of
the system behaviour, which is an alternative to the usual interpretation
involving Dicke eigenstates  (see for
example \cite{Rudolph:1995cn}).

\appendix

\section{Full Set Of Equations\label{chap:Full-Set-Of}}

\textcolor{black}{In this Appendix we present the equations of motion
for two atoms with the full $j_{u}$ internal structure, and arbitrary
orientation and laser polarisation. A convenient definition for the
atomic operators of the $i^{th}$ atom can be made in terms of the
irreducible tensor operators \cite{Edmonds}}

\textcolor{black}{
\begin{eqnarray}
T_{q}^{k}(ab)^{\left(i\right)} & = & \sum_{m_{a}}\sum_{m_{b}}(-1)^{j_{a}-m_{a}}\sqrt{2k+1}\nonumber \\
 &  & \left(\begin{array}{ccc}
j_{a} & k & j_{b}\\
-m_{a} & q & m_{b}
\end{array}\right)|j_{a}m_{a}\rangle^{\left(i\right)\left(i\right)}\langle j_{b}m{}_{b}|.\label{eq:SphericalTensor_defn}
\end{eqnarray}
For ease of notion we redefine these operators, including transforming
to slowly varying operators,}

\textcolor{black}{
\begin{equation}
n_{l}^{i}\equiv T_{0}^{0}(ll)^{(i)}=|j_{l}0\rangle^{(i)(i)}\langle j_{l}0|,
\end{equation}
}

\textcolor{black}{
\begin{equation}
d_{-}^{q,i}\equiv-e^{i\omega_{L}t}T_{q}^{1}(lu)^{(i)}=(-1)^{q}e^{i\omega_{L}t}|j_{l}0\rangle^{(i)(i)}\langle j_{u}-q|,
\end{equation}
}

\textcolor{black}{
\begin{equation}
d_{+}^{q,i}\equiv e^{-i\omega_{L}t}T_{q}^{1}(ul)^{(i)}=e^{-i\omega_{L}t}|j_{u}q\rangle^{(i)(i)}\langle j_{l}0|,
\end{equation}
}

\textcolor{black}{
\begin{eqnarray}
n_{\alpha\beta}^{i} & = & (-1)^{1+\alpha}\sum_{kq}\sqrt{2k+1}\left(\begin{array}{ccc}
1 & 1 & k\\
\alpha & -\beta & -q
\end{array}\right)T_{q}^{k}(uu)^{(i)}\nonumber \\
 & = & |j_{u}\alpha\rangle^{(i)(i)}\langle j_{u}\beta|,
\end{eqnarray}
}

noting these give the relation $\left(d_{+}^{q}\right)^{\dagger}=\left(-\right)^{q}d_{-}^{-q}$.

\subsection{\textcolor{black}{First Order Equations}}

\textcolor{black}{The first order equations of motion are } 
\begin{widetext}
\textcolor{black}{
\begin{align}
-i\frac{d}{dt}d_{-}^{q,i}= & \left[\Delta+\frac{i\gamma}{2}\right]d_{-}^{q,i}+\frac{(-1)^{q}}{2}\sum_{\alpha}\left((\mathbf{\Omega}_{L}^{i})_{\alpha}+2\sum_{\beta}(-1)^{\beta}g_{\beta\alpha}(\mathbf{R}_{ij})d_{-}^{-\beta,j}\right)\left(\delta_{-q\alpha}n_{l}^{i}-n_{\alpha(-q)}^{i}\right)+F(d_{-}^{q,i}),\label{eq:Tlu_full_ex}
\end{align}
}

\textcolor{black}{and}

\textcolor{black}{
\begin{align}
-i\frac{d}{dt}n_{q_{1}q_{2}}^{i}=i\gamma n_{q_{1}q_{2}}^{i} & -\frac{(-1)^{q_{2}}}{2}\left((\mathbf{\Omega}_{L}^{i})_{q_{1}}^{*}+2\sum_{\alpha}g_{\alpha q_{1}}^{*}(\mathbf{R}_{ij})d_{+}^{\alpha,j}\right)d_{-}^{-q_{2},i}\nonumber \\
 & +\frac{1}{2}\left(\left(\mathbf{\Omega}_{L}^{i}\right)_{q_{2}}+2\sum_{\alpha}(-1)^{\alpha}g_{\alpha q_{2}}(\mathbf{R}_{ij})d_{-}^{-\alpha,j}\right)d_{+}^{q_{1},i}+F(n_{q_{1}q_{2}}^{i}),
\end{align}
}

\textcolor{black}{where $j\neq i$ and the laser field $\mathbf{\Omega}_{L}^{i}=\frac{d}{\hbar\epsilon_{0}}\mathbf{D}^{{\rm ext}}\left(\mathbf{r}_{i}\right)=\Omega\boldsymbol{\varepsilon}_{L}e^{i\mathbf{k}_{L}\cdot{\bf r}_{i}}$
has polarisation $\boldsymbol{\varepsilon}_{L}$. The noise terms,
$F=F^{a}+F^{c}$, are given by}

\textcolor{black}{
\begin{equation}
F^{c}(d_{-}^{q,i})=(-1)^{q}\frac{\mathcal{D}}{\hbar}\left[e^{i\omega_{L}t}\sum_{\alpha}(-1)^{\alpha}(\mathbf{f}^{(c)}(\mathbf{r}_{i}))_{\alpha}\left(\delta_{q\alpha}n_{l}^{i}-n_{(-\alpha)(-q)}^{i}\right)\right],
\end{equation}
}

\textcolor{black}{
\begin{equation}
F^{a}(d_{-}^{q,i})=(-1)^{q}\frac{\mathcal{D}}{\hbar}\left[e^{i\omega_{L}t}\sum_{\alpha}(-1)^{\alpha}\left(\delta_{q\alpha}n_{l}^{i}-n_{(-\alpha)(-q)}^{i}\right)(\mathbf{f}^{(a)}(\mathbf{r}_{i}))_{\alpha}\right],
\end{equation}
}

\textcolor{black}{{} 
\begin{equation}
F^{c}(n_{q_{1}q_{2}}^{i})=(-1)^{q_{2}}\frac{\mathcal{D}}{\hbar}\left[e^{i\omega_{L}t}(\mathbf{f}^{(c)}(\mathbf{r}_{i}))_{-q_{2}}d_{+}^{q_{1},i}-e^{-i\omega_{L}t}(\mathbf{f}^{(c)}(\mathbf{r}_{i}))_{q_{1}}d_{-}^{-q_{2},i}\right],
\end{equation}
}

\textcolor{black}{and}

\textcolor{black}{{} 
\begin{equation}
F^{a}(n_{q_{1}q_{2}}^{i})=(-1)^{q_{2}}\frac{\mathcal{D}}{\hbar}\left[e^{i\omega_{L}t}d_{+}^{q_{1},i}(\mathbf{f}^{(a)}(\mathbf{r}_{i}))_{-q_{2}}-e^{-i\omega_{L}t}d_{-}^{-q_{2},i}(\mathbf{f}^{(a)}(\mathbf{r}_{i}))_{q_{1}}\right].
\end{equation}
} 
\end{widetext}

\textcolor{black}{where the operators $\mathbf{f}^{(c)}\left(\mathbf{r},t\right)$
and $\mathbf{f}^{(a)}\left(\mathbf{r},t\right)$ are the creation
and annihilations parts of the noise operator, $\mathbf{f}\left(\mathbf{r},t\right)=\mathbf{f}^{(a)}\left(\mathbf{r},t\right)+\mathbf{f}^{(c)}\left(\mathbf{r},t\right)$
which appears in Eq.(\ref{eq:Escatt_G}). They are given by}

\textcolor{black}{
\begin{eqnarray}
\mathbf{f}^{\left(a\right)}\left(\mathbf{r},t\right) & = & \left(\mathbf{f}^{\left(c\right)}\left(\mathbf{r},t\right)\right)^{\dagger}\label{eq:E_a_free}\\
 & = & i2\sqrt{3}\int{\rm d^{3}k}\mathcal{E}_{k}\sum_{\varepsilon\perp\mathbf{k}}\boldsymbol{\varepsilon}e^{i{\bf k}\cdot\mathbf{r}}e^{-i2\pi k\left(t-t_{0}\right)}a_{{\bf k}\boldsymbol{\varepsilon}}\left(t_{0}\right),\nonumber 
\end{eqnarray}
}

with $t_{0}$ the initial time and $\mathcal{E}_{k}\equiv\sqrt{\hbar k/2\epsilon_{0}\left(2\pi\right)^{2}}$.
The equations of motion are in normally ordered form 
\textcolor{black}{with respect to the field operators}, with all $\mathbf{f}_{n}^{\left(a\right)}\left(\mathbf{r},t\right)$
terms on the right and $\mathbf{f}_{n}^{\left(c\right)}\left(\mathbf{r},t\right)$
on the left, so that when we take the expectation value $\langle \rm{vac} |N^{a}|\rm{vac}\rangle=\langle \rm{vac}|N^{c}|\rm{vac}\rangle=0$
and these terms disappear. 

\subsection{\textcolor{black}{Second order Equations}}

The normally ordered second order equations are \textcolor{black}{obtained using the product rule to give}
\begin{widetext}
\textcolor{black}{
\begin{align}
-i\frac{d}{dt}d_{-}^{q_{1},i}d_{+}^{q_{2},j}= & i\gamma d_{-}^{q_{1},i}d_{+}^{q_{2},j}+\frac{(-1)^{q_{1}}}{2}(\mathbf{\mathbf{\Omega}}_{L}^{i})_{-q_{1}}n_{l}^{i}d_{+}^{q_{2},j}-\frac{1}{2}(\mathbf{\mathbf{\Omega}}_{L}^{j})_{q_{2}}^{*}d_{-}^{q_{1},i}n_{l}^{j}-(-1)^{q_{1}}\sum_{\alpha\beta}\left(g_{\beta\alpha}(\mathbf{R}_{ji})-g_{\alpha\beta}^{*}(\mathbf{R}_{ji})\right)n_{\alpha,(-q_{1})}^{i}n_{q_{2},\beta}^{j}\nonumber \\
 & -\frac{(-1)^{q_{1}}}{2}\sum_{\alpha}\left((\mathbf{\mathbf{\Omega}}_{L}^{i})_{\alpha}d_{+}^{q_{2},j}+2g_{\alpha q_{2}}^{*}(\mathbf{R}_{ji})n_{l}^{j}\right)n_{\alpha,(-q_{1})}^{i}+\frac{1}{2}\sum_{\alpha}\left((\mathbf{\mathbf{\Omega}}_{L}^{j})_{\alpha}^{*}d_{-}^{q_{1},i}+2(-1)^{q_{1}}g_{\alpha(-q_{1})}(\mathbf{R}_{ji})n_{l}^{i}\right)n_{q_{2},\alpha}^{j}\nonumber \\
 & +F(d_{-}^{q_{1},i}d_{+}^{q_{2},j}),
\end{align}
}

\textcolor{black}{{} 
\begin{align}
-i\frac{d}{dt}n_{q_{1}q_{2}}^{i}d_{-}^{\alpha,j}= & \left[\Delta+\frac{3i\gamma}{2}\right]n_{q_{1}q_{2}}^{i}d_{-}^{\alpha,j}+\frac{1}{2}\left((\mathbf{\Omega}_{L}^{i})_{q_{2}}d_{+}^{q_{1},i}-(-1)^{q_{2}}(\mathbf{\Omega}_{L}^{i})_{q_{1}}^{*}d_{-}^{-q_{2},i}\right)d_{-}^{\alpha,j}\\
 & +\frac{(-1)^{\alpha}}{2}\sum_{\beta}(\mathbf{\Omega}_{L}^{j})_{\beta}\left(\delta_{-\alpha\beta}n_{l}^{j}-n_{\beta(-\alpha)}^{j}\right)n_{q_{1}q_{2}}^{i}-(-1)^{\alpha+q_{2}}\sum_{\beta}g_{\beta q_{1}}^{*}(\mathbf{R}_{ij})d_{-}^{-q_{2},i}n_{\beta(-\alpha)}^{j}+F(n_{q_{1}q_{2}}^{i}d_{-}^{\alpha,j}),\nonumber 
\end{align}
}

\textcolor{black}{
\begin{align}
-i\frac{d}{dt}d_{-}^{q_{1},i}d_{-}^{q_{2},j}= & \left(2\Delta+i\gamma\right)d_{-}^{q_{1},i}d_{-}^{q_{2},j}+\frac{(-1)^{q_{1}}}{2}\sum_{\alpha}(\mathbf{\Omega}_{L}^{i})_{\alpha}\bigg(\delta_{\alpha-q_{1}}n_{l}^{i}-n_{\alpha-q_{1}}^{i}\bigg)d_{-}^{q_{2},j}\\
 & +\frac{(-1)^{q_{2}}}{2}\sum_{\alpha}(\mathbf{\Omega}_{L}^{j})_{\alpha}\bigg(\delta_{\alpha-q_{2}}n_{l}^{j}-n_{\alpha(-q_{2})}^{j}\bigg)d_{-}^{q_{1},i}+F(d_{-}^{q_{1},i}d_{-}^{q_{2},j}),\nonumber 
\end{align}
}

\textcolor{black}{and}

\textcolor{black}{
\begin{align}
-i\frac{d}{dt}n_{q_{1}q_{2}}^{i}n_{\alpha_{1}\alpha_{2}}^{j}= & 2i\gamma n_{q_{1}q_{2}}^{i}n_{\alpha_{1}\alpha_{2}}^{j}+\frac{1}{2}\left((\mathbf{\Omega}_{L}^{i})_{q_{2}}d_{+}^{q_{1},i}-(-1)^{q_{2}}(\mathbf{\Omega}_{L}^{i})_{q_{1}}^{*}d_{-}^{-q_{2},i}\right)n_{\alpha_{1}\alpha_{2}}^{j}\nonumber \\
 & +\frac{1}{2}\left((\mathbf{\Omega}_{L}^{j})_{\alpha_{2}}d_{+}^{\alpha_{1},j}-(-1)^{\alpha_{2}}(\mathbf{\Omega}_{L}^{j})_{\alpha_{1}}^{*}d_{-}^{-\alpha_{2},j}\right)n_{q_{1}q_{2}}^{i}+F(n_{q_{1}q_{2}}^{i}n_{\alpha_{1}\alpha_{2}}^{j}).\label{eq:TuuTuu_full_ex}
\end{align}
} 
\end{widetext}

\textcolor{black}{Given that  atomic operators from two different atoms commute at equal time} the noise terms for the second order equation's
for operators $A^{i}$ and $B^{j}$ can be written in terms of the
first order noise terms

\textcolor{black}{{} 
\begin{equation}
F^{c}(A^{i}B^{j})=F^{c}(A^{i})B^{j}+F^{c}(B^{j})A^{i},
\end{equation}
}

\textcolor{black}{{} 
\begin{equation}
F^{a}(A^{i}B^{j})=A^{j}F^{a}(B^{i})+B^{i}F^{a}(A^{j}).
\end{equation}
}

\textcolor{black}{Taking the expectation value, and choosing orientation
$\bk_{L}=k_{L}\mathbf{\hat{x}}$, $\mathbf{\Omega}_{L}=\Omega\mathbf{\hat{z}}$
and $\mathbf{R}\propto\mathbf{\hat{z}}$ or in the $x-y$ plane, these
equations reduce to two-state case. Setting $d_{-}^{i}=d_{-}^{0,i}$, $n_{u}^{i}=n_{00}^{i}$,
$n_{l}^{i}=1-n_{u}^{i}$ and $G(\mathbf{r})=g_{00}(\mathbf{r})$ 
gives Eq.(\ref{eq:dlu}-\ref{eq:nunu}).}


%

\end{document}